\documentclass[pra,10pt,twocolumn,superscriptaddress,showpacs]{revtex4-2}

\usepackage{epsfig}
\usepackage{amsmath}
\usepackage{graphicx}
\usepackage{graphics}
\usepackage{float}
\usepackage{amssymb}
\usepackage[usenames,dvipsnames]{color}
\usepackage{natbib}
\usepackage{epstopdf}
\usepackage{verbatim}
\usepackage{wrapfig}
\usepackage{mathtools}
\usepackage{dsfont}
\usepackage{breqn}

\usepackage{breqn}

\graphicspath{{./Figures/G}{./Figures/T}{./Figures/wc}}

\usepackage{bm}

\makeatletter
\let\cat@comma@active\@empty
\makeatother

\graphicspath{{Figures/}}

\newcommand{\bra}[1]{\left\langle #1\right|}
\newcommand{\ket}[1]{\left| #1\right\rangle}

\newcommand{\opav}[3]{\langle #1 | #2 | #3 \rangle}

\newcommand{\beq}{\begin{equation}}
\newcommand{\eeq}{\end{equation}}
\newcommand{\tr}{\text{Tr}}

\begin{document}

\title{Improving the estimation of environment parameters via initial probe-environment correlations}

\author{Hamza Ather}
\affiliation{School of Science \& Engineering, Lahore University of Management Sciences (LUMS), Opposite Sector U, D.H.A, Lahore 54792, Pakistan}

\author{Adam Zaman Chaudhry}
\email{adam.zaman@lums.edu.pk}
\affiliation{School of Science \& Engineering, Lahore University of Management Sciences (LUMS), Opposite Sector U, D.H.A, Lahore 54792, Pakistan}

\begin{abstract}

Small, controllable quantum systems, known as quantum probes, have been proposed to estimate various parameters characterizing complex systems such as the environments of quantum systems. These probes, prepared in some initial state, are allowed to interact with their environment, and subsequent measurements reveal information about different quantities characterizing the environment such as the system-environment coupling strength, the cutoff frequency, and the temperature. These estimates have generally been made by considering only the way that the probe undergoes decoherence. However, we show that information about the environment is also imprinted on the probe via the probe and environment correlations that exist before the probe state preparation. This information can then be used to improve our estimates for any environment. We apply this general result to the particular case of a two-level system probe undergoing pure dephasing, due to a harmonic oscillator environment, to show that a drastic increase in the quantum Fisher information, and hence the precision of our estimates, can indeed be obtained. We also consider applying periodic control pulses to the probe to show that with a combination of the two - the effect of the control pulses as well as the initial correlations - the quantum Fisher information can be increased by orders of magnitude. 

\end{abstract}

\maketitle

\section{\label{sec:level1}Introduction}

Realistic quantum systems are not isolated; their interactions with their surroundings generally need to be taken into account \cite{Gardinerbook, BPbook}. Studying such open quantum systems has been of great interest, driven by both fundamental aspects such as the quantum measurement problem as well as the practical implementation of futuristic quantum technologies. In order to investigate the effect of the environment on a quantum system, one needs to determine and characterize various properties of the environment. An effective way to tackle this complex problem is to use quantum probes \cite{ClaudiaPRA2014, Elliot2016,ViolaPRL2016,DarioPRA2016,Streif2016,Benedetti2017,CoscoPRA2017,SonePRA2017,ClaudiaPRA2020,WuPRA2020,Tamascelli2020,GiovannettiPhysRevR2020}. Quantum probes are small, simple, and easily controllable quantum systems which, prepared in a suitable initial state, are allowed to become correlated with the environment and consequently undergo decoherence. Suitable measurements are subsequently performed on the probe allowing us to infer the properties of the environment, with quantum estimation theory supplying the tools to quantify the precision of our estimates \cite{Helstrombook,Fujiwara2001,Monras2006,Monras2007,Paris2008QuantumEF,Genoni2011,Spagnolo2012,Pinel2013,ChaudhryNV2014,ChaudhryNV2015}. A key tool is the quantum Fisher information which determines the ultimate precision with which we estimate a given environment parameter. As dictated by the quantum Cramer-Rao bound, ideally we would like the quantum Fisher information to be as large as possible in order to obtain the best possible estimates.

To date, estimates of the environment parameters have been generally obtained by considering the decoherence of the probe once the probe and the environment have been prepared in a product state with the environment state being the thermal equilibrium state (recent work done in Refs.~\cite{WuPRA2020,Tamascelli2020} also includes dissipation). In other words, the correlations that develop after the probe state preparation are used to deduce the environment parameters. However, especially if the probe-environment interaction strength is strong, the probe and the environment will be correlated significantly before the probe state preparation as well. The role of such initial correlations in open quantum system dynamics has been widely investigated  \cite{HakimPRA1985, HaakePRA1985, Grabert1988, SmithPRA1990, GrabertPRE1997, PazPRA1997, LutzPRA2003, BanerjeePRE2003, vanKampen2004, BanPRA2009, HanggiPRL2009, UchiyamaPRA2010, TanimuraPRL2010, SmirnePRA2010, DajkaPRA2010, ZhangPRA2010,TanPRA2011, CKLeePRE2012,MorozovPRA2012, SeminPRA2012,  ChaudhryPRA2013a,ChaudhryPRA2013b,ChaudhryCJC2013,FanchiniSciRep2014,FanSciRep2015correlation,ChenPRA2016,VegaRMP2017,VegaPRA2017,ShibataJPhysA2017,CaoPRA2017,MehwishEurPhysJD2019}. Since these correlations are expected to impact the ensuing dynamics of the probe, we can expect the quantum Fisher information relevant to the estimation of a given environment parameter to be modified as well. Our central aim in this paper is to then consider a two-level system probe interacting with an environment whose parameters we want to estimate, taking into account the probe-environment correlations both before and after the probe state preparation. Hopefully, we can improve our estimates if the correlations before the probe state preparation are taken into account. 

We start our analysis by considering the general form of the time evolution of a two-level system probe undergoing only dephasing, provided that the probe is prepared in a pure state initially, without assuming any particular form of the environment. Ignoring relaxation effects due to the environment can be justified since relaxation timescales are typically much longer than dephasing timescales. We then calculate the quantum Fisher information for this general probe state and show that the Fisher information is the sum of two positive terms: one, the information obtained from the decoherence rate (which can also be modified by the initial correlations), and two, the information obtained from an effective level-shift of the quantum probe due to the initial correlations. In other words, the Fisher information now contains contributions due to the probe-environment correlations both before and after the probe state preparation. We then apply this general result to the particular case of a two-level system interacting with a collection of harmonic oscillators. The effect of the environment is encapsulated by the spectral density of the environment, which contains the coupling strength of the system and the environment, and the cutoff frequency which determines the correlation time of the environment. We aim to estimate the coupling strength, the cutoff frequency, as well as the temperature of the environment. We show that the effect of the initial correlations on the quantum Fisher information for these estimates can be drastic, especially for sub-Ohmic environments. Having determined the quantum Fisher information, we then discuss the measurements that can be used in practice in order to obtain these maximally precise estimates. Finally, we aim to further improve the quantum Fisher information by applying suitable control pulses to the system \cite{ViolaPRA1998,LloydPRL1999,Gordon2006,Yang2008,HansonScience2010,KhodjastehPRA2011,LiuFrontiers2011,ChaudhryPRA2013a,Tan2013,LidarPRL2018}. These pulses modulate both the decoherence rate as well as the effective level-shift. By considering these fields to be of pulse form, and then optimizing over the pulse interval as well as the interaction time of the probe with the environment, we find that, compared to the scenario where initial correlations are ignored and no pulses are applied, orders of magnitude improvement in the quantum Fisher information can be obtained. 

This paper is organized as follows. Section II explains the basic formalism that we will use throughout the paper. In Section III, we apply this general formalism to the parameters characterizing a harmonic oscillator environment. We then discuss in Section IV the measurements that practically need to be performed in order to maximize the Fisher information. In Section V, we aim to use suitable control pulses to optimize the quantum Fisher information. Finally, we summarize our findings in Section VI, with some technical details regarding the solution of our model given in the appendix.

\section{The Formalism}

Our objective is to estimate the various parameters, such as the temperature and the cutoff frequency, characterizing the environment of a quantum system. To this end, we use a two-level system coupled to the environment. We use this two-level system as a probe in the sense that from the dynamics of this probe, we can estimate the parameters characterizing the environment. In general, the initial pure state of the probe is 
\begin{gather} \label{eq:rho_0}
\rho(0)=\begin{pmatrix}
\cos^2(\frac{\theta_0}{2}) & \frac{1}{2}\sin \theta_0 e^{-i\phi_0} \\
\frac{1}{2}\sin \theta_0 e^{i\phi_0} & \sin^2(\frac{\theta_0}{2}),
\end{pmatrix}
\end{gather}
with $\theta_0$ and $\phi_0$ the usual angles on the Bloch sphere. As a result of the probe's interaction with the environment, the state of the probe evolves. Assuming that the environment causes predominantly decoherence of the probe state, the state of the probe at a later time $t$ can be written as \cite{Austin2020}
\begin{gather} \label{eq:rho_t_decoherence}
\rho(0)=\begin{pmatrix}
\cos^2(\frac{\theta_0}{2}) & \frac{1}{2}\sin \theta_0 e^{-i\Omega(t)} e^{-\Gamma(t)} \\
\frac{1}{2}\sin \theta_0 e^{i\Omega(t)}  e^{-\Gamma(t)} & \sin^2(\frac{\theta_0}{2})
\end{pmatrix}.
\end{gather}
Here $\Omega(t) = \omega_0 t + \phi_0 + \chi(t)$, where $\omega_0$ is the natural frequency of the probe, and $\chi(t)$ is the effect of a possible level shift induced by the environment. In studies performed to date, generally only the the information about the environment encoded in the function $\Gamma(t)$ has been considered. However, as we will show, information about the environment is also captured by the function $\chi(t)$, and using this additional information can improve our estimates regarding the environment parameters by orders of magnitude. 

To quantify the precision with which a general environment parameter $x$ can be estimated, we use the quantum Fisher information (QFI) given by \cite{Benedetti2017}
\beq \label{eq:QFI_generic}
H_Q(x)=\sum_{n=1}^2{\frac{(\partial_{x}\rho_n)^2}{\rho_n}}+2\sum_{n\neq m}{\frac{(\rho_n - \rho_m)^2}{\rho_n +\rho_m}|\langle{\epsilon_m}\ket{\partial_{x}\epsilon_n}|^2},
\eeq
where $\ket{\epsilon_n}$ is the $n^{\text{th}}$ eigenstate of our probe state, $\rho_n$ is the corresponding eigenvalue, and $\partial_x$ is just a short-hand notation for $\frac{\partial}{\partial x}$. To find the QFI given the general state at time $t$ for our probe, we first find that the eigenvalues of $\rho(t)$ are given by $\rho_{1} = \frac{1}{2}\left[1 + \mathcal{F}(t)\right]$ and $\rho_2 = \frac{1}{2}\left[1 - \mathcal{F}(t)\right]$, with $\mathcal{F}(t)=\sqrt{1+\sin^{2}\theta_{0}(e^{-2\Gamma(t)}-1)}$. The corresponding eigenstates are
\begin{align}
\ket{\epsilon_1(t)} &= \cos\left(\frac{\theta}{2}\right)\ket{0} + e^{i\Omega(t)}\sin\left(\frac{\theta}{2}\right)\ket{1}, \\
\ket{\epsilon_2(t)} &= \sin\left(\frac{\theta}{2}\right)\ket{0} - e^{i\Omega(t)}\cos\left(\frac{\theta}{2}\right)\ket{1},
\end{align}
where
\begin{align*}
    \sin\theta&= \mathcal{F}(t)^{-1}\sin\theta_{0}e^{-\Gamma(t)},\\
    \cos\theta&=\mathcal{F}(t)^{-1}\cos\theta_{0},\\
\end{align*}
and $\ket{0}$ and $\ket{1}$ are the usual eigenstates of $\sigma_z$ with $\sigma_z \ket{n} = (-1)^n \ket{n}$. We next find that 
$$ (\partial_x \rho_1)^2 = (\partial_x \rho_2)^2 = \frac{1}{4}[\mathcal{F}(t)]^{-2} \sin^4 \theta_0 e^{-4\Gamma} (\partial_x \Gamma)^2, $$
and 
$$ |\langle \epsilon_2(t) | \partial_x \epsilon_1(t)\rangle|^2 = \frac{1}{4}\left[(\partial_x \theta)^2 + \sin^2 \theta_0 (\partial_x \Omega)^2 \right], $$
with $|\langle \epsilon_2(t) | \partial_x \epsilon_1(t)\rangle|^2 = |\langle \epsilon_1(t) | \partial_x \epsilon_2(t)\rangle|^2$. Since $\tan \theta = (\tan \theta_0) e^{-\Gamma}$, we also find that 
$$ \partial_x \theta = -\sin \theta_0 \cos \theta_0 [\mathcal{F}(t)]^{-2} e^{-\Gamma} \partial_x \Gamma. $$
Putting all these results together, we are led to, after some algebraic manipulations, 
\begin{equation}
\label{generalQFI}
H_Q(x) = \frac{\sin^2 \theta_0}{e^{2\Gamma} - 1} \left(\frac{\partial \Gamma}{\partial x}\right)^2  + \sin^2 \theta_0 e^{-2\Gamma} \left(\frac{\partial \chi}{\partial x}\right)^2, 
\end{equation}
where we have used the fact that in $\Omega(t)$, only $\chi(t)$ can possibly depend on $x$. We emphasize that this is a totally general result for a two-level probe interacting with an arbitrary environment, irrespective of the probe-environment coupling. The only restriction is that we are restricting ourselves to pure dephasing. 

From the form of Eq.~\eqref{generalQFI}, we first conclude that independent of the interaction time of the probe with the environment, the maximum Fisher information is obtained for $\theta_0 = \pi/2$, as the contribution due to the decoherence factor $\Gamma(t)$ as well as the level shift $\chi(t)$ is maximized in this case. Consequently, we set $\theta_0 = \pi/2$ when we next apply our general formalism to the case of using a two-level probe to estimate the environment parameters for a harmonic oscillator environment. Before doing so, let us note that we can also easily generalize our formalism to the scenario where the initial probe state is not pure (see the appendix).

\section{Application for harmonic oscillator environment}

So far we have found the quantum Fisher information by considering an arbitrary environment. We now look at a particular environment. We consider a two-level system interacting with a bosonic environment. The system-environment Hamiltonian can be written as (we take $\hbar=1$ throughout) \cite{BPbook}
\beq \label{eq:hamiltonian}
H=\frac{\omega_0}{2}\sigma_z+\sum_{k}{\omega_k b_k^{\dagger} b_k}+\sum_{k}{}{\sigma_z(g_k b_k^{\dagger}+g_k^*b_k)}
\eeq
Here $\sigma_z$ is the standard Pauli spin matrix, $\omega_0$ is the energy splitting of the two-level system, $\omega_k$ is the frequency of the $k^{\text{th}}$ mode of the environment, $b_k$ and $b_k^{\dagger}$ are the standard lowering and raising operators, and $g_k$ is the interaction strength between the $k^{\text{th}}$ mode and the two-level system. The effect of the environment on the system is encapsulated by the spectral density of the environment $J(\omega)$. In this work, we assume that the spectral density is of the form \cite{Benedetti2017}
\begin{equation}\label{eq:spectral_density}
J(\omega)=G\frac{\omega^s}{\omega_c^{s-1}}e^{- \frac{\omega}{\omega_c}},
\end{equation}
where $\omega_c$ is the cutoff frequency, $G$ is the strength of the environment, and $s$ is the Ohmicity parameter. Besides the parameters appearing in the spectral density $J(\omega)$, that is, the cutoff frequency $\omega_c$ and the coupling strength $G$, we would also look to estimate the temperature $T$ of the environment using our general formalism. Before moving on, let us note for the sake of clarity of the future discussion that $s<1$ represents sub-Ohmic, $s>1$ super-Ohmic, and $s=1$ Ohmic spectral density.

\subsection{Estimating parameters when initial correlations are ignored}
The usual assumption in finding the dynamics of the two-level system probe is to assume that the probe state is independent of the environment \cite{BPbook}. In other words, the total initial system-environment state is taken to be $\rho_{\text{tot}} = \rho(0) \otimes \rho_B$, where $\rho(0)$ is the initial probe state, and $\rho_B = e^{-\beta H_B}/Z_B$, with $Z_B = \tr[e^{-\beta H_B}]$, is the environment thermal equilibrium state. Here $\beta = \frac{1}{k_B T}$ is the inverse temperature (we will set $k_B = 1$ throughout). With $\theta_0 = \pi/2$ and $\phi_0 = 0$, we find that the probe state at a later time $t$ is given by \cite{BPbook} 
\begin{gather*}
\rho(t)=\frac{1}{2}\begin{pmatrix}
1  &  e^{-\Gamma_{\text{uc}}}e^{-i\omega_0 t} \\ 
e^{i \omega_0 t} e^{-\Gamma_{\text{uc}}} & 1
\end{pmatrix},
\end{gather*}
where the decoherence factor (the `uc' denotes that this is for the uncorrelated initial system-environment state) is given by 
\begin{equation} \label{gammauc}
\Gamma_{\text{uc}}=\int_0^\infty G[1-\cos(\omega t)]\frac{\omega^{s-2}}{\omega_c^{s-1}}e^{-\frac{\omega}{\omega_c}}\coth\left(\frac{\omega}{2T}\right) d\omega.
\end{equation}
Notice that $\chi = 0$ in this case. Consequently, only the first term in Eq.~\eqref{generalQFI} contributes to the quantum Fisher information. Let us then look at the density matrix of the probe when initial correlations are taken into account to see how the quantum Fisher information changes.

\subsection{Effect of initial system-environment correlations on the quantum Fisher information}
Instead of considering the initial system environment state to be the product state $\rho(0) \otimes \rho_B$, we now consider a more realistic preparation of the initial probe state. We assume that the probe is in contact with the environment for a sufficiently long time such that equilibrium is reached. The total system-environment state is then proportional to $e^{-\beta H}$, where $H$ is the total system-environment Hamiltonian. This system-environment state takes into account any system-environment correlations. A projective measurement is then performed on the system at time $t = 0$ to prepare the desired probe state (with $\theta_0 = \pi/2$ and $\phi_0 = 0$). The system-environment state just after this measurement is then $\rho_{\text{tot}}(0) = \ket{\psi}\bra{\psi} \otimes \opav{\psi}{e^{-\beta H}}{\psi}/Z$, where $Z = \tr_B[\opav{\psi}{e^{-\beta H}}{\psi}]$. Even though this is a product state, the environment state is now different due to the previous correlations between the system and the environment. With this system-environment initial state, the state of the probe at time $t$ is \cite{MorozovPRA2012,ChaudhryPRA2013a,Austin2020}
\begin{align}
\rho(t)=\frac{1}{2}\begin{pmatrix}
1  &  e^{-\Gamma} e^{-i\chi(t)}e^{-i\omega_0 t} \\ 
e^{-\Gamma}e^{i\chi(t)}e^{i\omega_0 t} & 1
\end{pmatrix}, \label{rhowithcorrelation}
\end{align}
where 
\begin{equation*}
    \Gamma(t)=\Gamma_{\text{uc}}(t)+\Gamma_{\text{corr}}(t),
\end{equation*}
\begin{equation*}
  \Gamma_{\text{corr}}(t)=-\frac{1}{2}\ln\Bigg[1-\frac{\sin^2[\phi(t)]}{\cosh^2(\omega_0/2T)}\Bigg],  
\end{equation*}
\begin{equation*}
    \tan[\chi(t)]=\tanh(\omega_0/2T)\tan[\phi(t)],
\end{equation*}
and
\begin{equation*}
\phi(t)=\int_{0}^{\infty}\frac{J(\omega)}{\omega^2}\sin(\omega t) d\omega.
\end{equation*}
For completeness, we have sketched how to derive this result in the appendix. Note that $\Gamma_{\text{uc}}(t)$ is the same decoherence factor we had previously without taking into account system-environment correlations, $\Gamma_{\text{corr}}(t)$ is the correction to the decoherence factor due to the initial correlations, and $\phi(t)$ captures the effect of the time-dependent level shift induced due to the initial correlations. At zero temperature, this level shift is of particularly simple form, namely that $\chi(t) = \phi(t)$. Moreover,$\Gamma_{\text{corr}}(t) = 0$ at zero temperature. The important point to note is that now the information regarding the environment parameters is encoded in both the decoherence factor as well as the level shift. Consequently, now both terms in Eq.~\eqref{generalQFI} contribute, and, since both terms are guaranteed to be positive, we can expect that the quantum Fisher information will now increase. Before moving on to investigate this increase for the estimation of different environment parameters, let us note that approximately the same probe state $\rho(t)$ is obtained if, starting from the joint probe-environment equilibrium state, a unitary operation on the probe is performed to prepare the initial probe state at low temperatures \cite{MorozovPRA2012,ChaudhryPRA2013a,Austin2020}. Our results are then not just restricted to the case where a projective measurement is used to prepare the initial probe state. Moreover, if the initial probe state is prepared via a unitary operation at temperatures that are not very low, the initial probe state is mixed and a modified formalism (see the appendix) can be used instead to compute the quantum Fisher information.

\subsection{Estimating the cutoff frequency $\omega_c$}

While estimating $\omega_c$, we assume $T=0$ to simplify our results. Application of Eq.~\eqref{generalQFI} requires us to find $\frac{\partial \Gamma}{\partial \omega_c}$ and $\frac{\partial \chi}{\partial \omega_c}$. For our model, both of these can be found analytically. We find that for $s \neq 1$, 
\begin{multline*}
    \frac{\partial \Gamma}{\partial \omega_c}=\frac{G}{\omega_c}\Bigg[(1-s)\Bar{\Gamma}[s-1] + \Bar{\Gamma}[s] \Bigg]\\ -\frac{G(1-s)}{2\omega_c}\Bigg[\frac{1}{(1+i\omega_c t)^{s+1}}+\frac{1}{(1-i\omega_c t)^{s+1}} \Bigg]\Bar{\Gamma}[s-1] \\ -\frac{G}{2\omega_c}\Bigg[\frac{1}{(1+i\omega_c t)^{s}}+\frac{1}{(1-i\omega_c t)^{s}} \Bigg]\Bar{\Gamma}[s]
\end{multline*}
while for $s=1$,
\begin{equation*}
    \frac{\partial \Gamma}{\partial \omega_c}=\frac{G\omega_c t^2}{1+(\omega_c t)^2}.
\end{equation*}
Here $\Bar{\Gamma}[x]$ denotes the usual gamma function. Since for low temperature $\chi(t)=\phi(t)$, we also find that 
\begin{multline*}
    \frac{\partial \phi}{\partial\omega_c}=\frac{G}{2i\omega_c}\Bigg[\bigg[ \frac{1}{(1-i\omega_c t)^{s-1}}-\frac{1}{(1+i\omega_c t)^{s-1}}\bigg](1-s)\Bar{\Gamma}[s-1] \\+ \bigg[ \frac{1}{(1-i\omega_c t)^{s}}-\frac{1}{(1+i\omega_c t)^{s}}\bigg]\Bar{\Gamma}[s] \Bigg],
\end{multline*}
for $s \neq 1$, while for $s = 1$, 
\begin{equation*}
    \frac{\partial \phi}{\partial\omega_c}=\frac{G t}{1+(\omega_c t)^2}.
\end{equation*}

\begin{figure}[b!]
    \centering
\includegraphics[width=9cm, height =5cm]{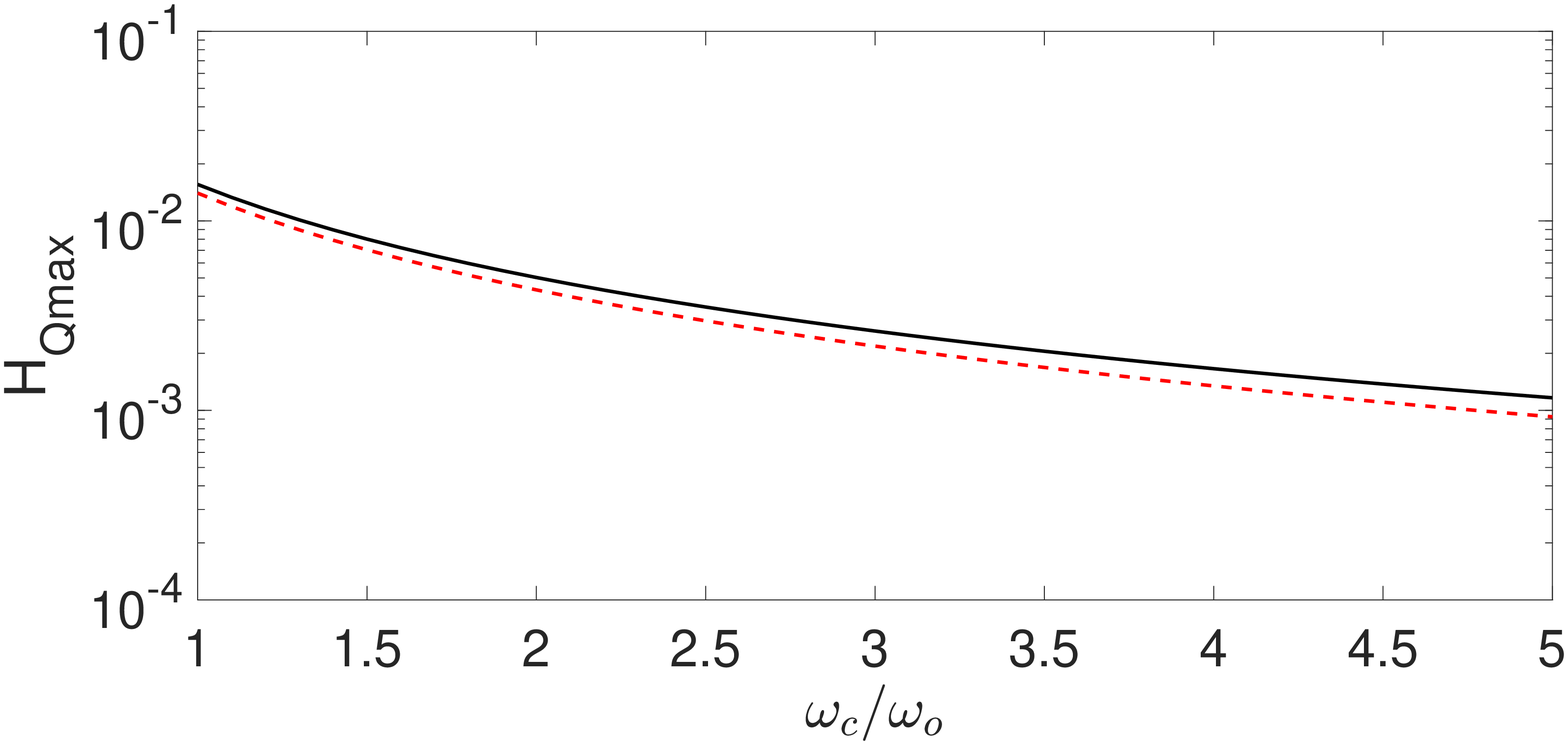}
    \caption{{The optimized quantum Fisher information as a function of the cutoff frequency.  The solid line refers to the Fisher information we get when taking into account the initial system-environment correlations while the dashed line is the Fisher information without including the initial system-environment correlations. Throughout, we are working in dimensionless units with $\hbar = 1$, and we have set $\omega_0 = 1$. Here, $s=0.5$, $G=0.01$ at $T=0$. }}
    \label{fig1}  
\end{figure}

\begin{figure}[h!]
    \centering
\includegraphics[width=9cm, height =5cm]{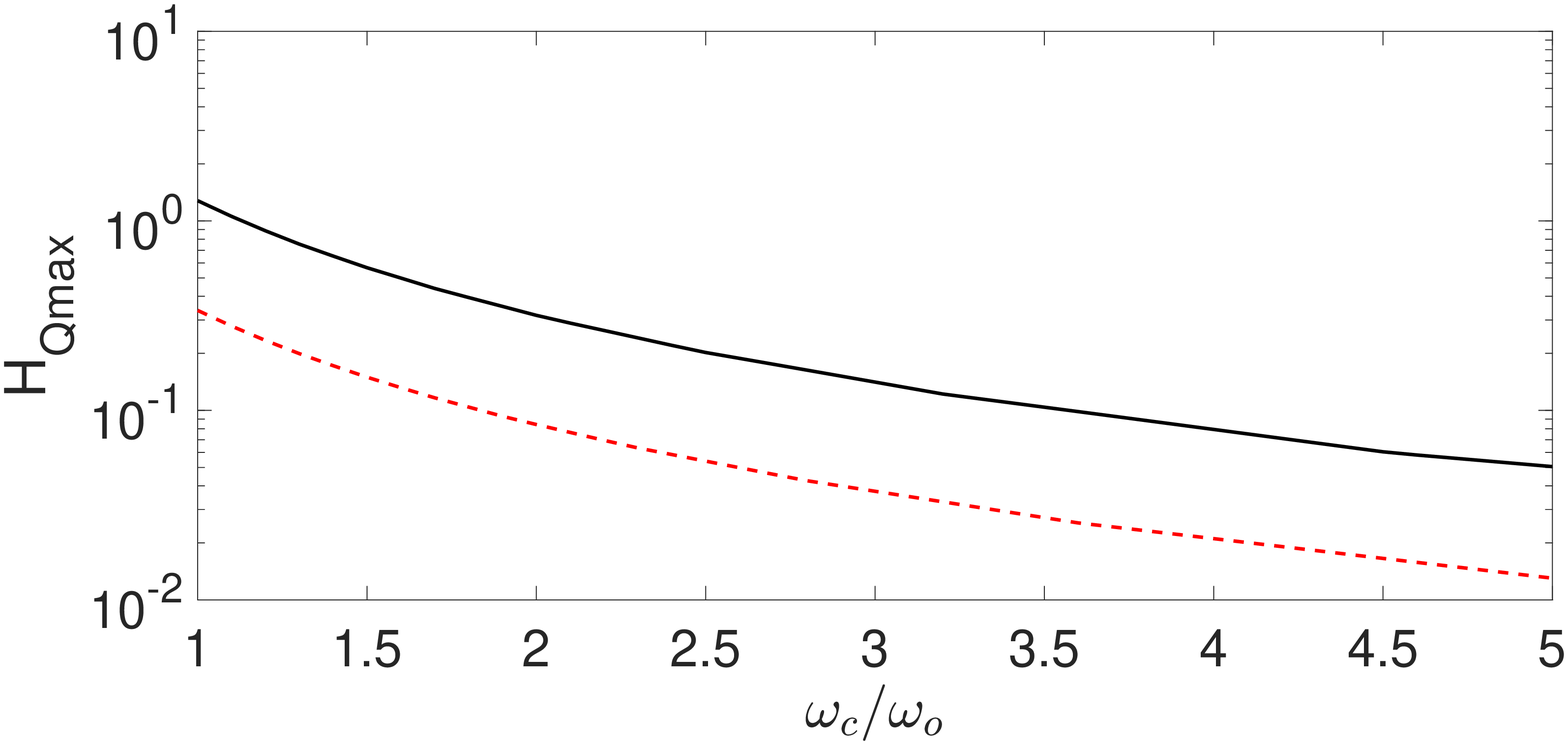}
    \caption{Same as Fig.~\ref{fig1}, except that now $G = 1$.}
    \label{fig2}
\end{figure}

\begin{figure}[h]
    \centering
\includegraphics[width=9cm, height =5cm]{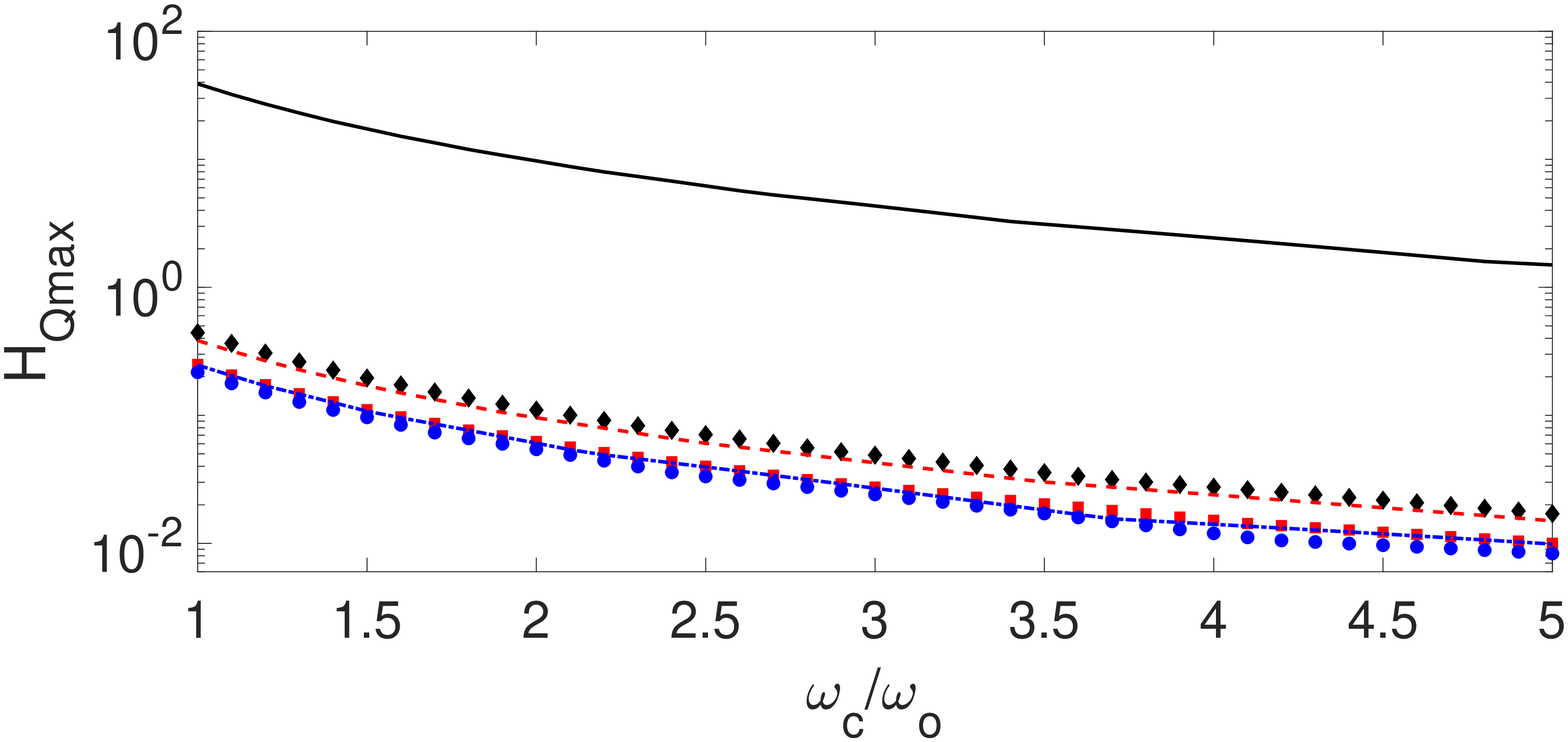}
    \caption{{The optimal value of the quantum Fisher information when estimating $\omega_c$. The solid, black curve shows the Fisher information for $s = 0.1$ when initial correlations are taken into account, while the black diamonds show the Fisher information for the same value of $s$ but ignoring the initial correlations. The dashed, red curve shows the Fisher information for $s = 2$ and taking into account the initial correlations, while the red squares do not include the initial correlations. Finally, the dot-dashed blue curve illustrates the behavior of the Fisher information for $s = 2$ with initial correlations included, and the blue circles ignore the initial correlations. Here we have used $G = 1$ and $T = 0$.}}
    \label{fig3}
\end{figure}

With these expressions in hand, we can evaluate the quantum Fisher information. However, the quantum Fisher information so evaluated will be function of the interaction time $t$ between the probe and the environment. To find the maximum possible Fisher information, we need to maximize over the interaction time to find the optimal time of interaction $t_{\text{opt}}$. It is this maximum value of the Fisher information that we consider to illustrate the effect of the initial correlations on our estimate of the cutoff frequency. In Fig.~\ref{fig1}, we have shown this optimized quantum Fisher information as a function of the cutoff frequency in the weak system-environment coupling regime for a sub-Ohmic environment with $s = 0.5$. The solid, black curve, which takes into account the initial probe-environment correlations, and the dashed, red curve, which is obtained by not considering the initial correlations, are close to each other, thereby signifying that the role played by the initial correlations in this case is small. This is expected, since in this weak coupling regime, the role played by the initial correlations is expected to be small. Moreover, the quantum Fisher information is seen to decrease as the cutoff frequency increases; this is simply a manifestation of the fact that as the cutoff frequency increases, more modes of the environment contribute to the decoherence of the probe. As the coupling strength is increased (see Fig.~\ref{fig2}), it is clear that much better estimates of the cutoff frequency can be obtained if the initial correlations are taken into account. We have also investigated the effect of the initial correlations as the Ohmicity parameter $s$ is varied. As illustrated in Fig.~\ref{fig3}, the biggest advantage of the initial correlations is obtained for sub-Ohmic environments, where an improvement of more than an order of magnitude can be obtained; for Ohmic and super-Ohmic environments, the increase in quantum Fisher information is more modest. This is expected since initial correlations are expected to play a larger role with sub-Ohmic environments with their longer memory time. 

\subsection{Estimating the coupling strength $G$}
\begin{figure}
    \centering
    \includegraphics[width=9cm, height =5cm]{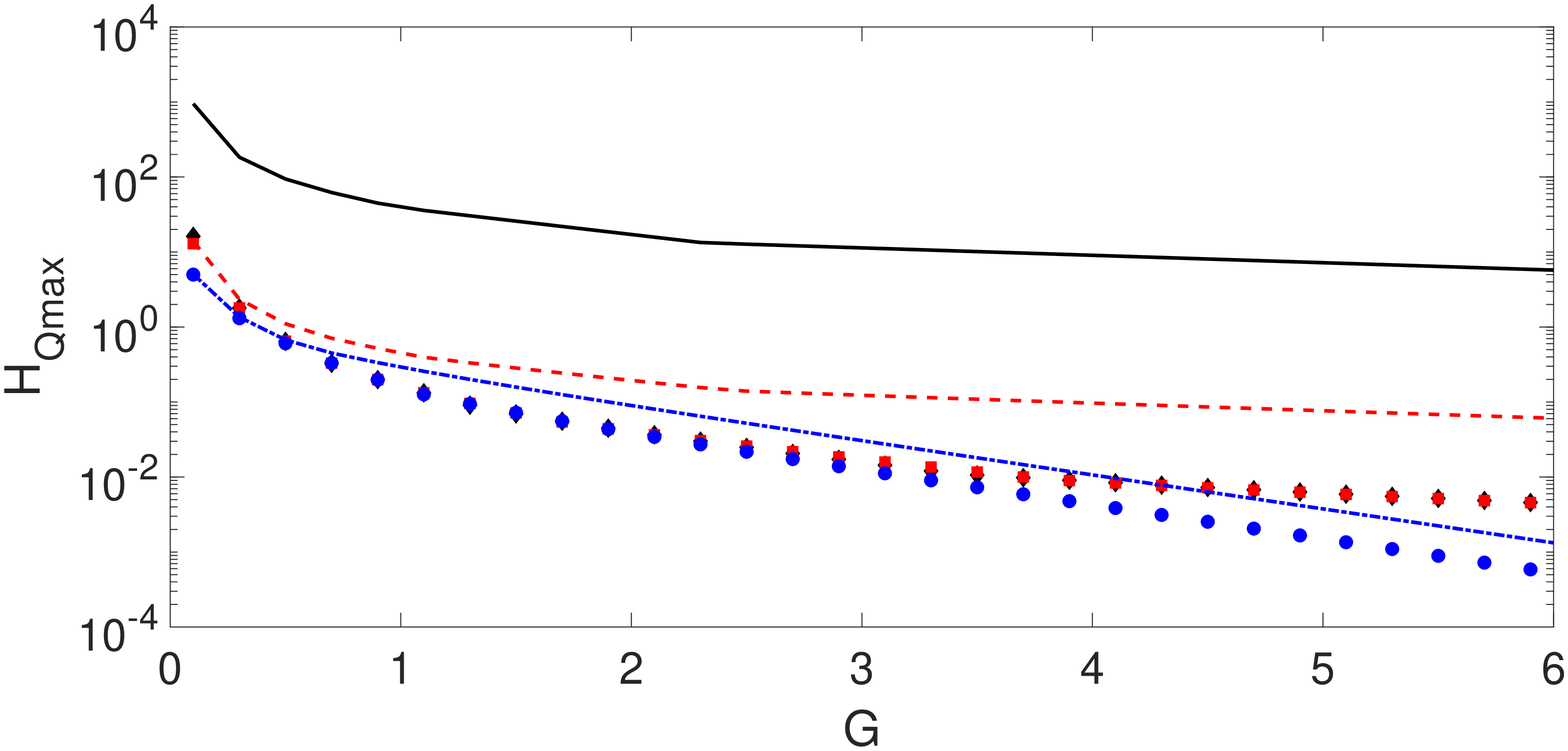}
    \caption{{Optimal value of the quantum Fisher information when estimating the coupling strength $G$.  The solid, black curve shows the Fisher information for $s = 0.1$, taking into account the initial correlations, while the black diamonds illustrate the behavior of the Fisher information for the same $s$ but without taking into account the initial correlations. The red-dashed curve shows the Fisher information for $s = 1$ and with initial correlations, while the red squares do not include the initial correlations. Similarly, the dot-dashed, blue curve corresponds to $s = 3$ with initial correlations accounted for, and the blue circles do not take into account the initial correlations. As before, we are using dimensionless units with $\hbar = 1$ and we have set $\omega_0 = 1$. We have used $\omega_c=5$ and $T = 0$.}}
    \label{fig4combined}
\end{figure}

Having estimated the cutoff frequency, we now estimate the system-environment coupling strength parameter $G$. To evaluate the quantum Fisher information now, we need $\frac{\partial \Gamma}{\partial G}$ and $\frac{\partial \chi}{\partial G}$. We now find that, assuming $T = 0$,
\begin{equation*}
	\frac{\partial \Gamma}{\partial G}=
	\begin{cases}
	 \Bigg[1-\frac{1}{2}\bigg[\frac{1}{(1+i\omega_c t)^{s-1}}+\frac{1}{(1-i\omega_c t)^{s-1}}\bigg] \Bigg]\Bar{\Gamma}[s-1] & s\neq  1  \\
    \frac{1}{2}\log(1+(\omega_c t)^2) & s=1,
	\end{cases}
\end{equation*}
and 
\begin{equation*}
    \frac{\partial \chi}{\partial G}=
    \begin{cases}
       \frac{1}{2i}\Bigg[\frac{1}{(1-i\omega_c t)^{s-1}}-\frac{1}{(1+i\omega_c t)^{s-1}} \Bigg]\Bar{\Gamma}[s-1] & s\neq 1\\
       \tan^{-1}(\omega_c t) & s=1.
    \end{cases}  
\end{equation*}
These expressions are used in Eq.~\eqref{generalQFI}, and the optimized Fisher information is then investigated as a function of the coupling strength for different Ohmicity parameters, with and without initial correlations. The results are shown in Fig.~\ref{fig4combined}. It is clear that once again, for sub-Ohmic environments at least, we obtain a drastic improvement of around two orders of magnitude in the quantum Fisher information, and hence our estimate regarding the coupling strength, if the initial correlations are accounted for. With Ohmic and super-Ohmic environments, the improvement, while undoubtedly present, is less impressive.

\subsection{Estimating the temperature $T$}
\begin{figure}
    \centering
    \includegraphics[width=9cm, height =5cm]{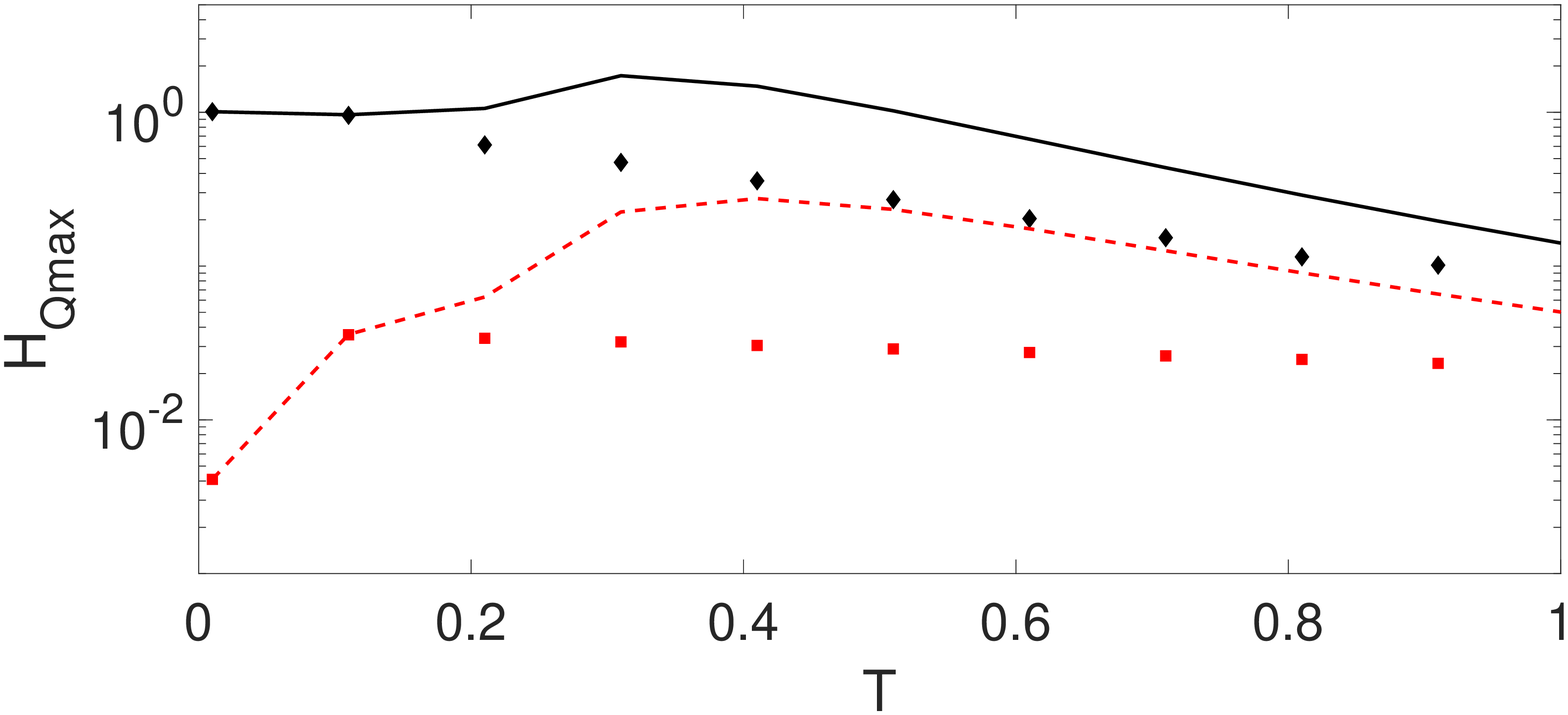}
    \caption{{Behavior of the maximum quantum Fisher information for a sub-Ohmic environment ($s = 0.1$) taking into account the initial correlations (solid, black curve) and without initial correlations (black diamonds). The dashed, red curve shows the Fisher information for an Ohmic environment, including the initial correlations, while the red squares do not include the initial correlations. As always, we have set $\hbar = 1$ with $\omega_0 = 1$. Also, $G=1$ and $\omega_c=5$.}}
    \label{figT}
\end{figure}

There are considerable differences with the two previous scenarios when we try to estimate the temperature. Since we can no longer take $T=0$, not only does $\Gamma_{\text{uc}}$ have a hyperbolic tangent factor, but now $\Gamma_{\text{corr}}$ is also no longer equal to zero. Moreover, $\chi(t)$ is no longer simply equal to $\phi(t)$. This means we have to use the full expressions in given below Eq.~\eqref{rhowithcorrelation}. We now find that 
\begin{equation*}
    \frac{\partial \Gamma_{\text{uc}}}{\partial T}=\int_0^\infty \frac{G[1-\cos(\omega t)]}{2T^2}\frac{\omega^{s-1}}{\omega_c^{s-1}}e^{-\frac{\omega}{\omega_c}}\cosh^2\left(\frac{\omega}{2T}\right) d\omega,
\end{equation*}
and 
\begin{equation*}
    \frac{\partial \Gamma_{\text{corr}}}{\partial T}=\frac{\omega_0^2}{2}\Bigg[\frac{\tanh(\frac{\omega_0}{2T})\sin^2[\phi(t)]}{T^2(\cosh^2(\frac{\omega_0}{2T})-\sin^2[\phi(t)])} \Bigg].
\end{equation*}
Also,
\begin{equation*}
    \frac{\partial \chi}{\partial T}=-\frac{\omega_0^2\tan[\phi(t)][1-\tanh^2\left(\frac{\omega_0}{2T}\right)]}{2\lbrace 1+ \tanh\left(\frac{\omega_0}{2T}\right)\tan[\phi(t)])\rbrace^2 T^2},
\end{equation*}
with 
\begin{equation*}
    \phi(t)=\begin{cases}
       \frac{G}{2i}\Bigg[\frac{1}{(1-i\omega_c t)^{s-1}}-\frac{1}{(1+i\omega_c t)^{s-1}} \Bigg]\Bar{\Gamma}[s-1] & s\neq 1\\
       G \tan^{-1}(\omega_c t) & s=1.
    \end{cases}     
\end{equation*}
We evaluate the quantum Fisher information using these expressions and then, as before, optimize it over the time taken for the probe and the environment to interact. The maximum Fisher information as a function of the temperature to be estimated is shown in Fig.~\ref{figT}. It is clear that once again, the initial correlations can substantially help us in improving our estimates, and, as before, the advantage is most prominent for sub-Ohmic environments. 

\section{Optimal Measurement}

\begin{figure}
    \centering
    \includegraphics[width=9cm, height =5cm]{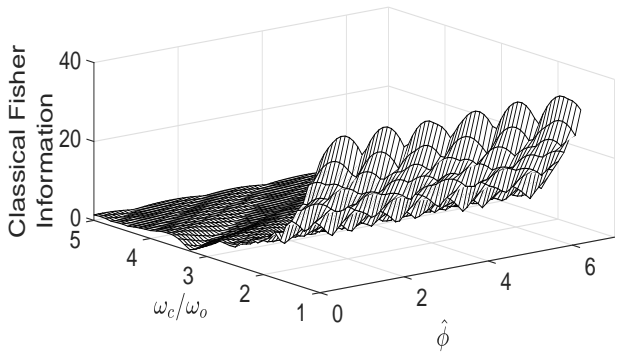}
    \caption{{The Fisher information $F_c(\omega_c)$, optimized over the interaction time $t$, obtained for a range of $\hat{\phi}$ from 0 to $2\pi$ when estimating $\omega_c$. Here, we have $G=1$, $T=0$, and $s=0.1$.}}
    \label{figclassicalfisher}
\end{figure}

\begin{figure}
    \centering
    \includegraphics[width=9cm, height =5cm]{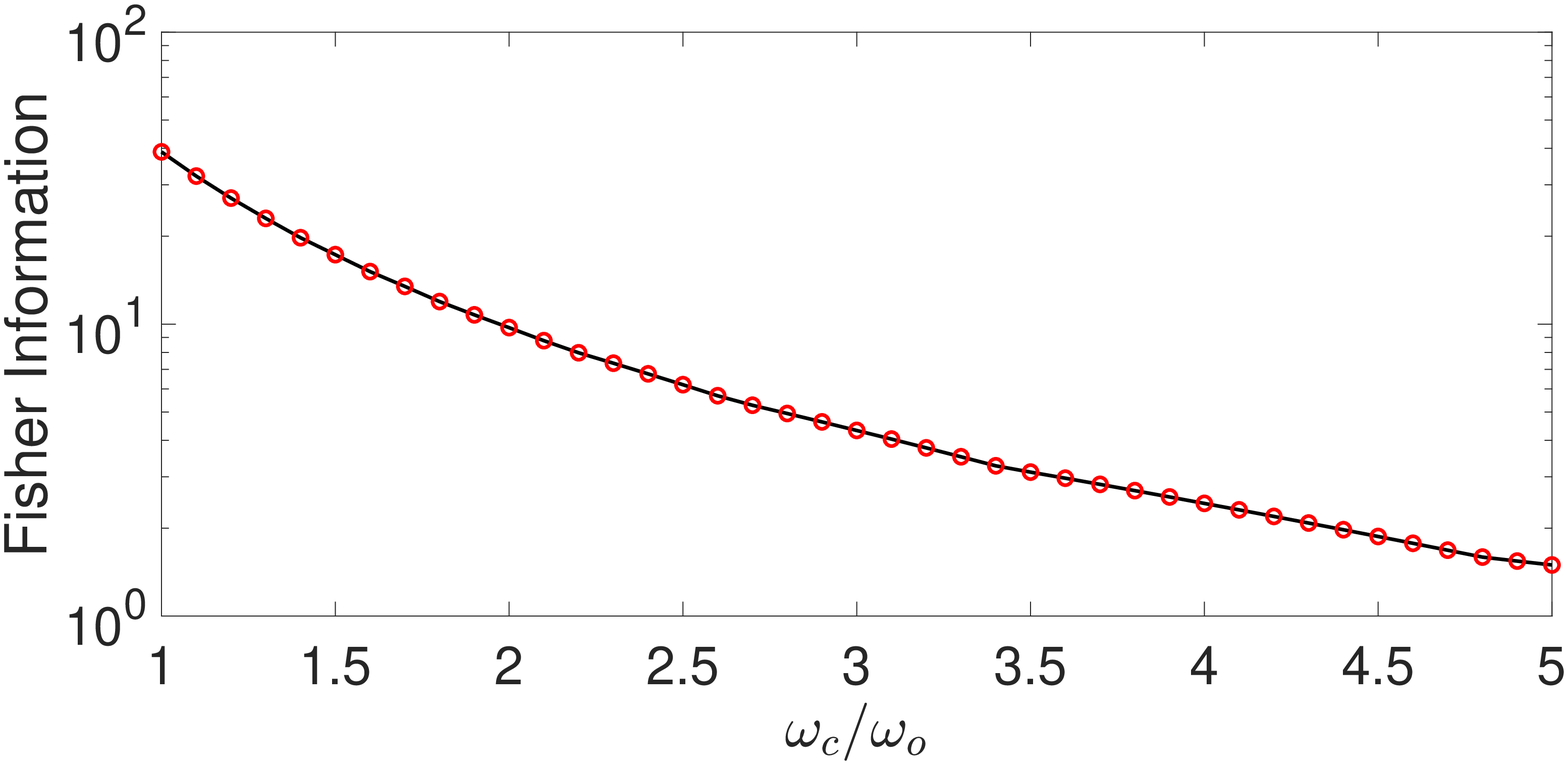}
    \caption{{Comparison of $F_c(\omega_c)$, optimized over the interaction time $t$, and the quantum Fisher information for the estimation of $\omega_c$. The solid line represents $F_c(\omega_c)$, while the circles represent the quantum Fisher information. Here, we have used $G=1$, $T=0$ and $s=0.1$.}}
    \label{figcomparisonoptomega}
\end{figure}

\begin{figure}
    \centering
    \includegraphics[width=9cm, height =5cm]{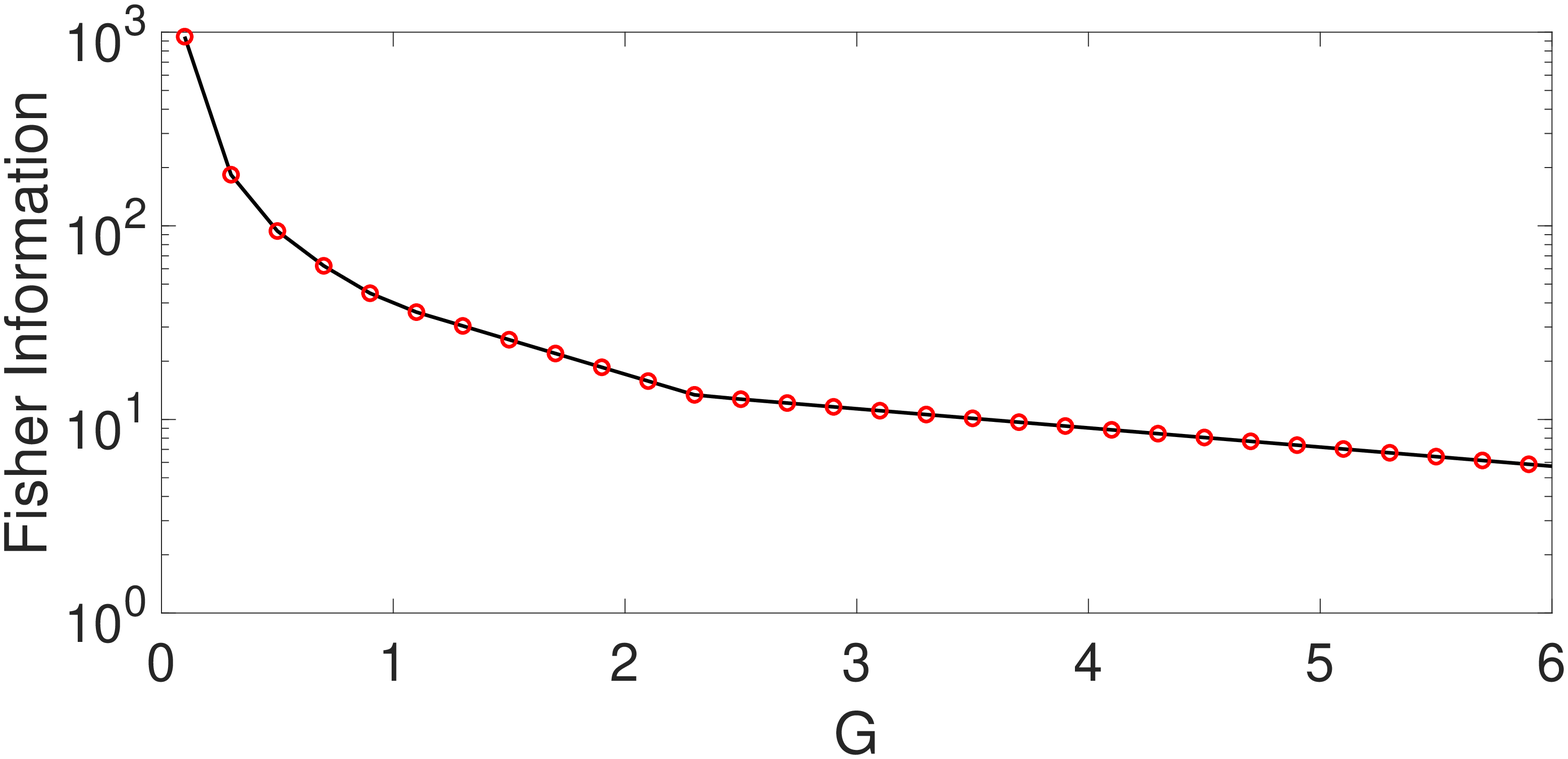}
    \caption{{Same as Fig.~\ref{figcomparisonoptomega}, except that we are now estimating the coupling strength $G$, and $\omega_c = 5$.}}
    \label{figcomparisonoptG}
\end{figure}

\begin{figure}
    \centering
    \includegraphics[width=9cm, height =5cm]{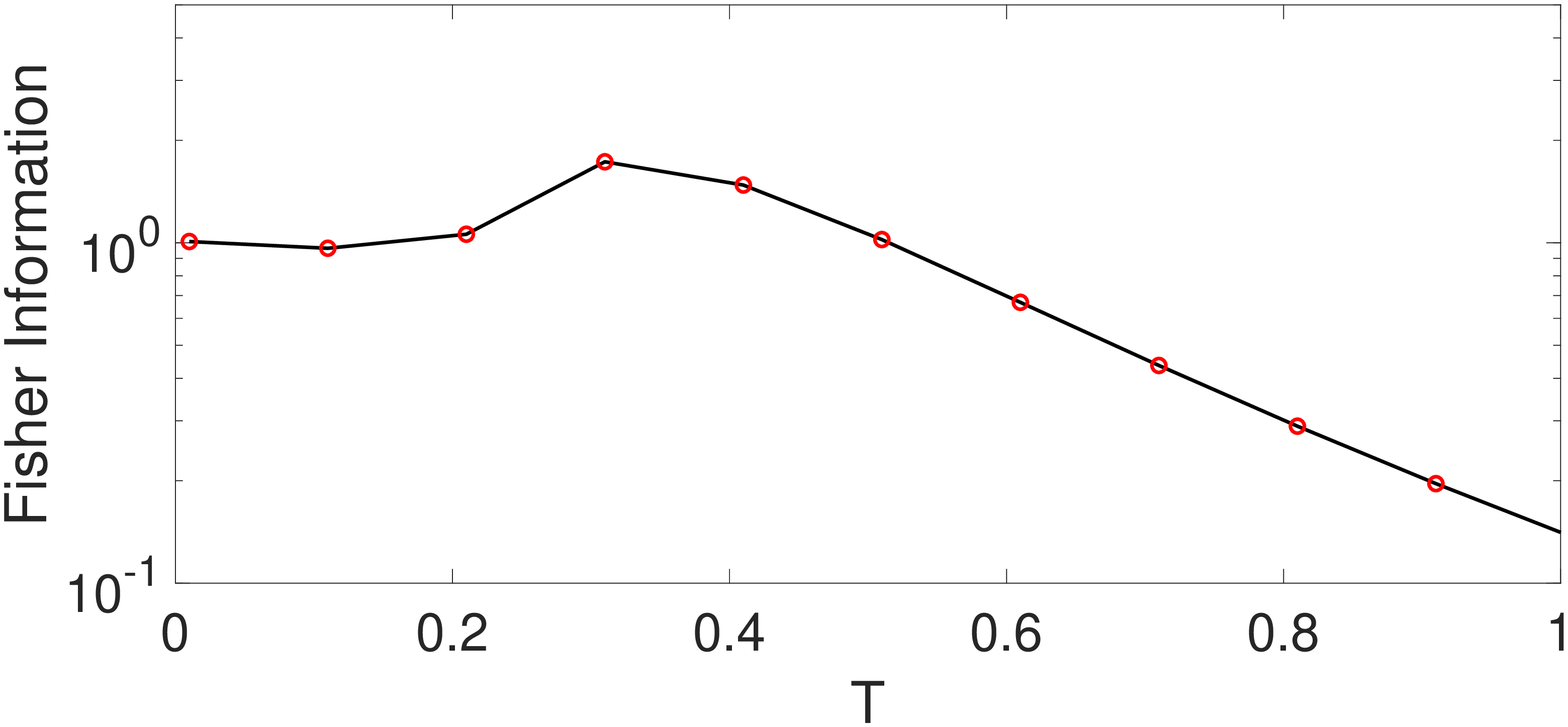}
    \caption{Same as Fig.~\ref{figcomparisonoptomega}, except that we are now estimating the temperature $T$, and $\omega_c = 5$.}
    \label{figcomparisonoptT}
\end{figure}

Until now, we have been discussing the quantum Fisher information, which, given the probe state, dictates how well we can estimate a given parameter if the most optimal measurement is performed on the probe. Having found that the quantum Fisher information is substantially enhanced due to the initial correlations, we now turn to finding the optimal measurement that needs to be performed on the probe state. In estimation theory, if the probe gives us measurement result $y$ when we are trying to estimate the unknown measurement $x$, the measurement can be characterized by the likelihood function $P(y|x)$. Our aim is to then minimize the mean-square error in our estimate. This is minimized by maximizing the Fisher information, since the Cramer-Rao bound says that 
\beq \label{eq:CRB}
\text{err}(x)\geq\frac{1}{F_c(x)} 
\eeq
where $F_c(x)$ is the Fisher information defined as
\begin{equation}
F_c(x)=\int\Bigg(\frac{\partial^2 \ln[P(y|x)]}{\partial x^2}\Bigg)P(y|x)dy.
\label{classicalFisher}
\end{equation}
The Fisher information given in Eq.~\eqref{classicalFisher} obviously depends on the measurements that we perform on the probe. The quantum Fisher information is the maximum $F_c(x)$ over all possible measurements; therefore, for the optimal measurement, $H_Q(x)$ and $F_c(x)$ will coincide. We can use this to infer the measurements that need to be performed on the probe. Since our probe state, for the optimal quantum Fisher information, lies in the equatorial plane of the Bloch sphere, our first guess is that the optimum measurements are projective measurements with the projectors given by  
\begin{equation*}
\hat{P_1}=\ket{\Psi_1}\bra{\Psi_1}, \quad \hat{P_2}=\ket{\Psi_2}\bra{\Psi_2},
\end{equation*}
with
\beq \label{eq:measuremet_vectors}
\ket{\Psi_{1,2}}=\frac{1}{\sqrt{2}}\ket{0} \pm \frac{e^{i\hat{\phi}}}{\sqrt{2}}\ket{1}.
\eeq
Using these projectors and the probe state given by Eq.~\eqref{rhowithcorrelation}, we find that 
\begin{equation}
F_c(x)=\frac{\bigg[ \cos(\omega_0 t+\chi - \hat{\phi})\big(\frac{\partial\Gamma}{\partial x}\big) + \sin(\omega_0 t+\chi - \hat{\phi})\big(\frac{\partial\chi}{\partial x}\big) \bigg]^2}{e^{2\Gamma}-\cos^2(\omega_0 t+\chi - \hat{\phi})}.
\label{eqclassicalfisher}
\end{equation}
This Fisher information depends on the parameter we are estimating $x$, the angle $\hat{\phi}$, and the interaction time $t$. For example, suppose we are estimating the cutoff frequency $\omega_c$. We can maximize the Fisher information $F_c(\omega_c)$ over time to find the optimal interaction time. With this interaction time, we can consider the behavior of $F_c(\omega_c)$ as a function of $\omega_c$ and the angle $\hat{\phi}$. Results are illustrated in Fig.~\ref{figclassicalfisher}, which clearly shows the dependence of $F_c(\omega_c)$ on $\hat{\phi}$. Our goal then is to maximize the Fisher information with respect to the angle $\hat{\phi}$. Observing Eq.~\eqref{eqclassicalfisher}, it is clear that if $\chi = 0$, then $\hat{\phi} = \omega_0 t$ is the optimal value of $\hat{\phi}$, and $F_c(x)$ reduces to the quantum Fisher information without taking into account the initial correlations. However, if $\chi \neq 0$, a simple, albeit tedious, optimization leads us to the conclusion that 
\beq\label{phihatopt}
\hat{\phi}=\omega_0 t+\chi -\text{tan}^{-1}\left[\frac{\left( \frac{\partial\chi}{\partial x}\right)\left( e^{2\Gamma}-1\right)}{\left(\frac{\partial\Gamma}{\partial x}\right)e^{2\Gamma}} \right]
\eeq
is the angle $\hat{\phi}$ that gives us the maximum Fisher information $F_c(x)$. Using this value of $\hat{\phi}$ in $F_c(x)$ [see Eq.~\eqref{eqclassicalfisher}], we find that $F_c(x)$ so found reduces to the quantum Fisher information given in Eq.~\eqref{generalQFI}. Very importantly, as is clear from Eq.~\eqref{phihatopt}, the optimum measurements that need to be performed depend on the initial correlations. We have numerically checked these analytical results for the optimal measurements in Figures \ref{figcomparisonoptomega}, \ref{figcomparisonoptG} and \ref{figcomparisonoptT} which show a comparison between the classical Fisher information with $\hat{\phi}$ given by Eq.~\eqref{phihatopt} and the quantum Fisher information for the estimation of the cutoff frequency $\omega_c$, the coupling strength $G$, and the temperature $T$ respectively. Since the solid lines and the circles overlap, our measurements are indeed optimal.

\section{Using control pulses to further improve the quantum Fisher information}

\begin{figure}
    \centering
    \includegraphics[width=9cm, height =5cm]{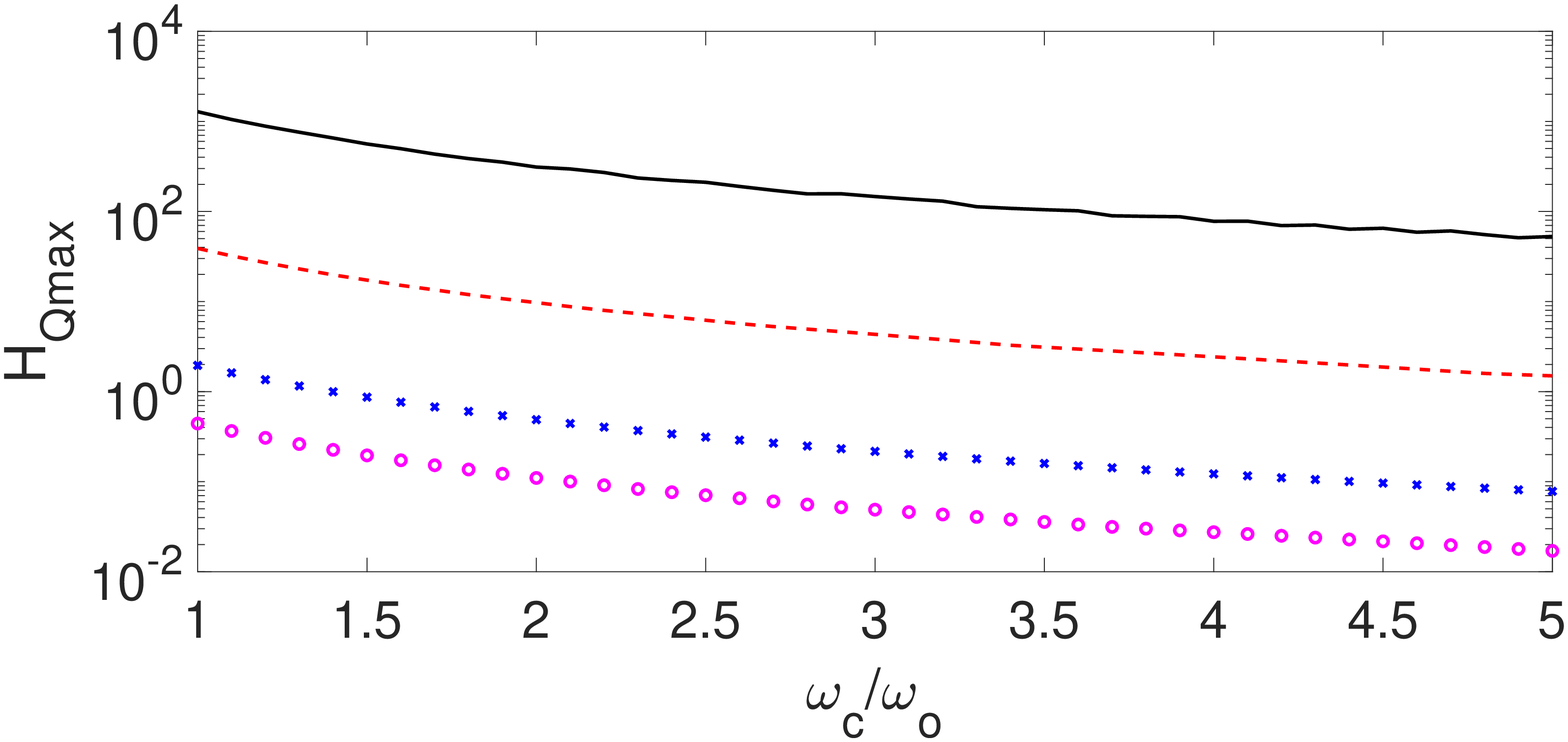}
    \caption{{Optimum value of the quantum Fisher information for the estimation of $\omega_c$ for a sub-Ohmic environment with $s = 0.1$. The dashed, red line refers to the QFI we get while taking into account the initial system-environment correlations, without control pulses, while the $\circ$ markers indicate the quantum Fisher information we get without taking these correlations into account and without control pulses. We then apply periodic pulses in both scenarios. We get the blue crosses when we do not take into account the initial correlations but apply control pulses and we get the solid, black line when we take into account the initial system-environment correlations with control pulses applied as well. As always, we have set $\hbar = 1$ with $\omega_0 = 1$. Here we have used $G=1$ and $T=0$. }}
    \label{figomegacwithpulses}
\end{figure}

\begin{figure}
    \centering
    \includegraphics[width=9cm, height =5cm]{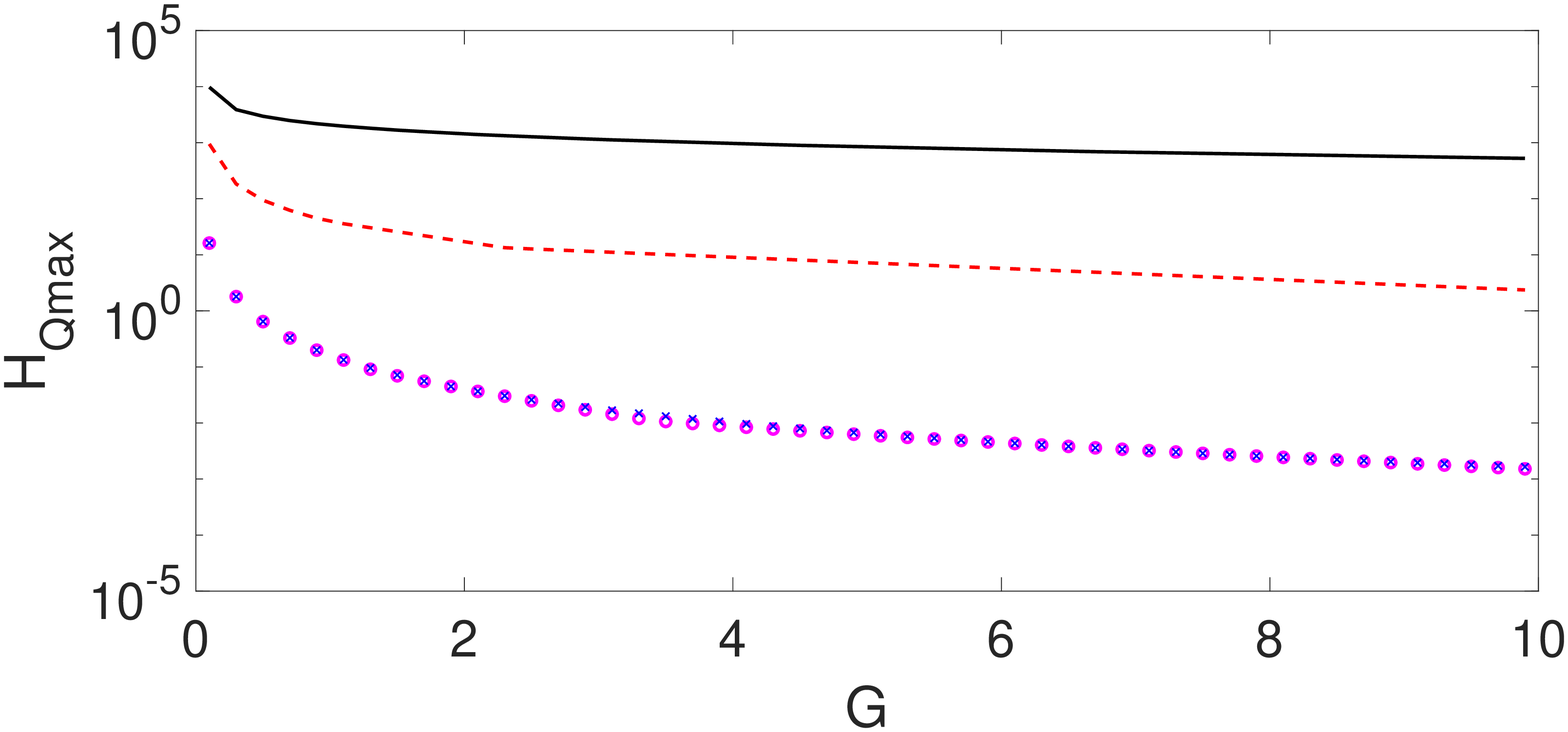}
    \caption{{Same as Fig.~\ref{figomegacwithpulses}, except that we are now estimating the coupling strength $G$, with $\omega_c = 5$.}}
    \label{figGwithpulses}
\end{figure}

As we have seen, the interaction of the probe with its environment allows us to estimate the environment parameters. To further improve our estimates, we can think of applying suitable control fields to the probe while it is interacting with the environment \cite{ViolaPRL2016,MullerSciRep2018}. Such control fields can effectively modulate the interaction of the probe with the environment, thereby possibly allowing the probe to collect more information about the environment before the probe completely decoheres. For our simple two-level system undergoing decoherence due to a harmonic oscillator environment, we will consider the periodic application of $\pi$-pulses to the probe. These $\pi$-pulses effectively keep on switching the sign of the probe-environment interaction Hamiltonian. To take this into account mathematically, it is useful to introduce the function 
\begin{equation} \label{eq:pulses_function}
\gamma(s,n)=\sum_{j=0}^{n}(-1)^j \Theta(s- t_j)\Theta(t_{j+1}-s),
\end{equation} 
where $n$ is the number of pulses, $s\in[0,t]$, $t_j = \frac{jt}{n + 1}$, and $\Theta(x)$ is defined as
\begin{equation*}
\Theta(x)= \begin{cases} 
+1 & ; \quad x > 0 \\
-1 & ; \quad x < 0
\end{cases}. 
\end{equation*}
The modulated system-environment Hamiltonian can be written as 
\begin{equation} \label{eq:hamiltonian_pulses}
H=\frac{\omega_0}{2}\sigma_z + \sum_{k}\omega_k b_{k}^{\dagger}b_k + \sum_{k}\gamma(s,n)\sigma_z(g_kb_{k}^{\dagger}+g_{k}^{*}b_k)
\end{equation}
This modified Hamiltonian leads to a different decoherence factor. In particular, we now find that 
\begin{equation} \label{eq:gamma_uc_pulses}
\Gamma_{\text{uc}}(t)=\int_{0}^{\infty}J(\omega)F_n(\omega,t)d\omega
\end{equation}
where
\begin{equation*}
F_n(\omega,t)=\tan^2\Big(\frac{\omega t}{2n+2}\Big)\Big(\frac{1+(-1)^n \cos(\omega t)}{\omega^2}\Big).
\end{equation*}
Also, we now find that 
\begin{equation} \label{eq:correlation_pulses}
    \phi(t)=\frac{1}{2}\int_0^{\infty}G\omega^{s-2}\omega_c^{1-s}e^{-\frac{\omega}{\omega_c}}M_n(\omega,t) d\omega,
\end{equation}
where
\begin{equation*}
    M_n(\omega,t)=(-1)^{n+1}\sin(\omega t)+2\sum_{j=1}^n(-1)^j \sin\Big(\frac{j \omega t}{n+1}\Big).
\end{equation*}
$\Gamma_{\text{corr}}(t)$ and $\chi(t)$ are then modified accordingly since 
\begin{equation*}
  \Gamma_{\text{corr}}(t)=-\frac{1}{2}\ln\Bigg[1-\frac{\sin^2[\phi(t)]}{\cosh^2(\omega_0/2T)}\Bigg],  
\end{equation*}
\begin{equation*}
    \tan[\chi(t)]=\tanh(\omega_0/2T)\tan[\phi(t)],
\end{equation*}
are still valid. With these modified factors at hand, $\Gamma(t)$ and $\chi(t)$ can easily be calculated. We then look to again estimate $\omega_c$, $G$, and $T$. Our aim will be to optimize the quantum Fisher information by modifying the total interaction time of the probe with the environment as well as the number of pulses to be applied during this interval. Results for the estimation of the cutoff frequency $\omega_c$ are shown in Fig.~\ref{figomegacwithpulses} for a sub-Ohmic environment, since this is the scenario in which the initial correlations play the most significant role. The magenta circles show the optimized quantum Fisher information if initial correlations are not considered, and no pulses are applied either. The dashed, red curve shows the drastic improvement in the quantum Fisher information when initial correlations are taken into account. With applied pulses, the improvement is even more drastic (see the solid, black curve). In short, orders of magnitude improvement in the Fisher information is obtained if initial correlations are taken into account and control pulses are also applied. Similar results hold when the coupling strength $G$ is estimated (see Fig.~\ref{figGwithpulses}). However, in this case, we find that the improvement in the estimation of the temperature using pulses is much more modest; taking the initial correlations into account plays a far more important role.

\begin{figure}
    \centering
    \includegraphics[width=9cm, height =5cm]{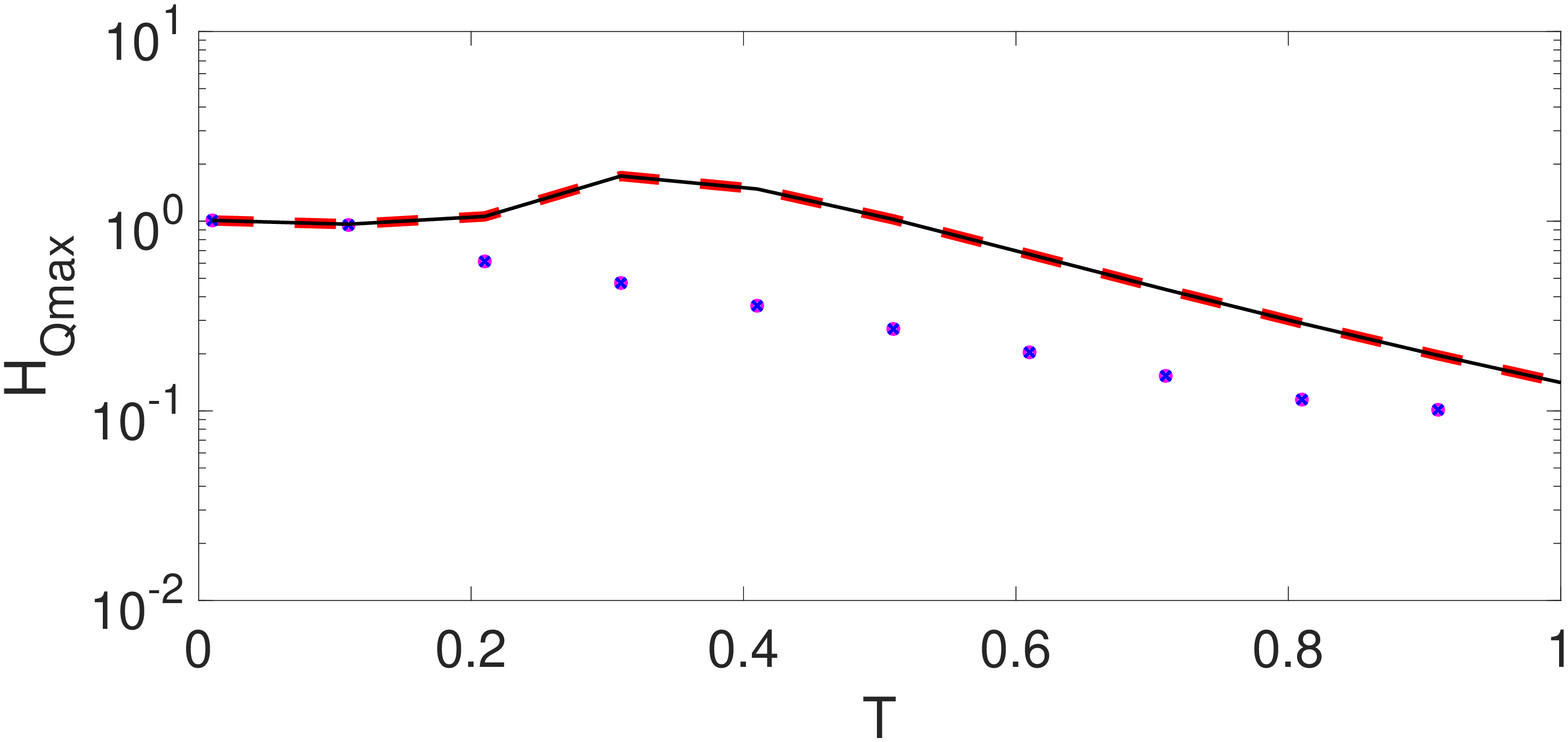}
    \caption{{Same as Fig.~\ref{figomegacwithpulses}, except that we are now estimating the temperature $T$, with $\omega_c = 5$.}}
    \label{figTwithpulses}
\end{figure}

\section{Conclusion}
In conclusion, we have found an exact expression for the quantum Fisher information for a two-level quantum probe that is used to estimate an arbitrary environment parameter, valid if the probe is undergoing pure dephasing. We make no assumption in this expression regarding the form of the environment or the probe-environment interaction strength. This expression is able to take into account the effect of any probe-environment correlations that existed before the probe state preparation, and shows a possible increase in the quantum Fisher information and hence an increase in the precision of our estimates. We then applied this general expression to the case of a probe interacting with an environment consisting of a collection of harmonic oscillators. We showed that, especially for relatively strong probe-environment coupling strength with sub-Ohmic environments, the initial correlations between the probe and the environment can greatly improve the quantum Fisher information. In this sense, both the correlations that develop between the probe and the environment after the probe state preparation as well as the correlations that exist before the probe state preparation are useful for the estimation of the environment. We then discussed which measurements on the probe actually need to be performed for the harmonic oscillator environment to obtain the most precise estimates. Importantly, these optimal measurements also depend on the initial correlations. Finally, we applied control pulses to the probe so as to affect the way that the probe gets correlated with the environment after the probe state preparation. We found that by applying suitable control pulses, the quantum Fisher information, as a result of both the initial correlations and the control pulses, is increased by orders of magnitude. Our results should be very useful for quantum noise sensing.

\section*{acknowledgements}
The authors acknowledge support from the LUMS FIF Grant. A.~Z.~C. is also grateful for support from HEC under grant No 5917/Punjab/NRPU/R\&D/HEC/2016. Support from the National Center for Nanoscience and Nanotechnology is also acknowledged.

\appendix

\section{The quantum Fisher information when the initial probe state is mixed}

If the initial probe state is mixed (as would be the case if the initial probe state is prepared via a unitary operation performed on the probe), the state of the probe can be written as \cite{Austin2020}
\begin{gather} \label{eq:rho_mixed}
\rho(0)=\begin{pmatrix}
\cos^2(\frac{\widetilde{\theta}_0}{2}) & \frac{1}{2}\sin \widetilde{\theta}_0 e^{-i\phi_0} e^{-\Gamma_0} \\
\frac{1}{2}\sin \widetilde{\theta}_0 e^{i\phi_0}  e^{-\Gamma_0} & \sin^2(\frac{\widetilde{\theta}_0}{2}),
\end{pmatrix}
\end{gather}
where $\Gamma_0 > 0$ takes into account the fact that the initial probe state is mixed. Note that $\widetilde{\theta}_0$ is not necessarily a Bloch angle here. It follows that at time $t$,
\begin{gather} \label{eq:rho_t_mixed}
\rho(t)=\begin{pmatrix}
\cos^2(\frac{\widetilde{\theta}_0}{2}) & \frac{1}{2}\sin \widetilde{\theta}_0 e^{-i\Omega(t)} e^{-\Gamma(t)} \\
\frac{1}{2}\sin \widetilde{\theta}_0 e^{i\Omega(t)}  e^{-\Gamma(t)} & \sin^2(\frac{\widetilde{\theta}_0}{2}),
\end{pmatrix}
\end{gather}
with $\Omega(t)=\omega_0 t+\chi(t)+\phi_0$  and $\Gamma(t)=\Gamma_{\text{uc}}(t)+\Gamma_0+\Gamma_{\text{corr}}(t)$. The eigenvalues are 
$$\rho_{1,2}=\frac{1}{2}\left[1\pm\widetilde{\mathcal{F}}(t)\right],$$
with 
$$\widetilde{\mathcal{F}}(t)=\sqrt{1+\text{sin}^2\widetilde{\theta}_0(e^{-2\Gamma}-1)},$$
and the corresponding eigenvectors are
\begin{align}
\ket{\epsilon_1(t)} &= \cos\left(\frac{\widetilde{\theta}}{2}\right)\ket{0} + e^{i\Omega(t)}\sin\left(\frac{\widetilde{\theta}}{2}\right)\ket{1}, \\
\ket{\epsilon_2(t)} &= \sin\left(\frac{\widetilde{\theta}}{2}\right)\ket{0} - e^{i\Omega(t)}\cos\left(\frac{\widetilde{\theta}}{2}\right)\ket{1},
\end{align}
where
\begin{align*}
    \sin\widetilde{\theta}&= \mathcal{F}(t)^{-1}\sin\widetilde{\theta}_{0}e^{-\Gamma(t)},\\
    \cos\widetilde{\theta}&=\mathcal{F}(t)^{-1}\cos\widetilde{\theta}_{0}.\\
\end{align*}
We then find the quantum Fisher information to be
\begin{equation}\label{generalQFImixed}
H_Q(x) = \frac{\sin^2 \widetilde{\theta}_0}{e^{2\Gamma} - 1} \left(\frac{\partial \Gamma}{\partial x}\right)^2  + \sin^2 \widetilde{\theta}_0 e^{-2\Gamma} \left(\frac{\partial \chi}{\partial x}\right)^2.
\end{equation}
\\
This expressions is analogous to what has been found for the case where the initial probe state is pure.

\section{Derivation of time evolution of density matrix cross-terms for system state preparation by projective measurement}
For our pure dephasing model, only the off-diagonal element of the density matrix, $[\rho(t)]_{10}$ need to be found as the diagonal elements do not change with time and the other off-diagonal is related by complex conjugation. The total system-environment Hamiltonian is \cite{MorozovPRA2012,ChaudhryPRA2013a} 
\begin{align*}
&H_S = \frac{\omega_0}{2} \sigma_z, \; H_B = \sum_k \omega_k b_k^\dagger b_k, \\
&H_{SB} = \sigma_{z}\sum_k (g_k^* b_k + g_k b_k^\dagger).
\end{align*}
First, we transform to the interaction picture to obtain
\begin{align}
H_I(t) &= e^{i(H_S + H_B)t} H_{\text{SB}} e^{-i(H_S + H_B)t}, \notag \\
&= \sigma_z \sum_k (g_k^* b_k e^{-i\omega_k t} + g_k b_k^\dagger e^{i\omega_k t} ).
\end{align}
We next find the time evolution operator $U_I(t)$ corresponding to $H_I(t)$ using the Magnus expansion to be
\begin{equation}
U_I(t) = \exp \lbrace \sigma_z \sum_k [b_k^\dagger \alpha_k(t) - b_k \alpha_k^*(t)]/2\rbrace,
\end{equation}
with 
$$\alpha_k(t) = \frac{2g_k (1 - e^{i\omega_k t})}{\omega_k}.$$
The total unitary time-evolution operator is then $U(t) = e^{-i\omega_0 \sigma_z t/2} U_I(t)$. Now, the off-diagonal element can be written as 
\begin{equation}
[\rho(t)]_{10}=\text{Tr}_{\text{S,B}}[U(t)\rho(0)U^{\dagger}(t)\ket{0}\bra{1}].
\end{equation} 
This depends on the initial system-environment state. If we consider the initial total state to be of the form $\rho_{\text{tot}} = \rho(0) \otimes e^{-\beta H_B}/Z_B$, with $\rho$ the initial state of the probe and $Z_B=\text{Tr}_B[e^{-\beta H_B}]$, then we can evaluate the trace over the system and the environment to obtain \cite{BPbook}
\begin{equation}
[\rho(t)]_{10}=[\rho(0)]_{10}e^{i\omega_0 t}e^{-\Gamma_{\text{uc}}(t)}
\end{equation}
This is the usual result. However, we aim to include the initial correlations. To this end, we suppose that the probe and the environment have reached the joint equilibrium state proportional to $e^{-\beta H}$. The quantum probe state is then prepared at $ t = 0$. The joint probe-environment state is then written as 
\begin{equation}
\rho_{\text{tot}}(0)=\frac{\Omega e^{-\beta H}\Omega^ {\dagger}}{Z},
\end{equation}
where $Z$ is the normalization factor and $\Omega$ is an operator which describes the probe preparation procedure - it can be a projection operator or a unitary operator. This initial state can then be used to find the time-dependence of the off-diagonal elements. Using a unitary transformation that `displaces' the environment harmonic oscillators, it is possible to show that now 
\begin{align}
[\rho(t)]_{10} &= [\rho(0)]_{10} e^{i\omega_0 t}   e^{-\Gamma_{\text{uc}}(t)} X(t),
\end{align}
with 
$$X(t) = \dfrac{\sum_l  \opav{l}{\Omega^\dagger \ket{0} \bra{1} \Omega}{l} e^{-i(-1)^l\phi(t)} e^{-\beta \omega_0 (-1)^l/2} }{\sum_l \opav{l}{\Omega^\dagger \ket{0} \bra{1} \Omega}{l} e^{-\beta \omega_0 (-1)^l/2} }$$
Assuming that $\Omega$ is a projection operator, that is, $\Omega = \ket{\psi}\bra{\psi}$, we can further simplify and write $X(t)$ in polar form to obtain 
\begin{equation}
[\rho(t)]_{10}=[\rho(0)]_{10}e^{i\omega_0 t}e^{i\chi(t)}e^{-\Gamma(t)},
\end{equation}
where $\Gamma(t)$ and $\chi(t)$ are defined below Eq.~\eqref{rhowithcorrelation} in the main text. We get the same result at low temperatures if $\Omega$ is instead a unitary operator which tries to prepare the same initial probe state.


\begin{thebibliography}{66}%
\makeatletter
\providecommand \@ifxundefined [1]{%
 \@ifx{#1\undefined}
}%
\providecommand \@ifnum [1]{%
 \ifnum #1\expandafter \@firstoftwo
 \else \expandafter \@secondoftwo
 \fi
}%
\providecommand \@ifx [1]{%
 \ifx #1\expandafter \@firstoftwo
 \else \expandafter \@secondoftwo
 \fi
}%
\providecommand \natexlab [1]{#1}%
\providecommand \enquote  [1]{``#1''}%
\providecommand \bibnamefont  [1]{#1}%
\providecommand \bibfnamefont [1]{#1}%
\providecommand \citenamefont [1]{#1}%
\providecommand \href@noop [0]{\@secondoftwo}%
\providecommand \href [0]{\begingroup \@sanitize@url \@href}%
\providecommand \@href[1]{\@@startlink{#1}\@@href}%
\providecommand \@@href[1]{\endgroup#1\@@endlink}%
\providecommand \@sanitize@url [0]{\catcode `\\12\catcode `\$12\catcode
  `\&12\catcode `\#12\catcode `\^12\catcode `\_12\catcode `\%12\relax}%
\providecommand \@@startlink[1]{}%
\providecommand \@@endlink[0]{}%
\providecommand \url  [0]{\begingroup\@sanitize@url \@url }%
\providecommand \@url [1]{\endgroup\@href {#1}{\urlprefix }}%
\providecommand \urlprefix  [0]{URL }%
\providecommand \Eprint [0]{\href }%
\providecommand \doibase [0]{https://doi.org/}%
\providecommand \selectlanguage [0]{\@gobble}%
\providecommand \bibinfo  [0]{\@secondoftwo}%
\providecommand \bibfield  [0]{\@secondoftwo}%
\providecommand \translation [1]{[#1]}%
\providecommand \BibitemOpen [0]{}%
\providecommand \bibitemStop [0]{}%
\providecommand \bibitemNoStop [0]{.\EOS\space}%
\providecommand \EOS [0]{\spacefactor3000\relax}%
\providecommand \BibitemShut  [1]{\csname bibitem#1\endcsname}%
\let\auto@bib@innerbib\@empty
\bibitem [{\citenamefont {Gardiner}\ and\ \citenamefont
  {Zoller}(2004)}]{Gardinerbook}%
  \BibitemOpen
  \bibfield  {author} {\bibinfo {author} {\bibfnamefont {C.~W.}\ \bibnamefont
  {Gardiner}}\ and\ \bibinfo {author} {\bibfnamefont {P.}~\bibnamefont
  {Zoller}},\ }\href@noop {} {\emph {\bibinfo {title} {Quantum noise}}}\
  (\bibinfo  {publisher} {Springer},\ \bibinfo {address} {Berlin},\ \bibinfo
  {year} {2004})\BibitemShut {NoStop}%
\bibitem [{\citenamefont {Breuer}\ and\ \citenamefont
  {Petruccione}(2007)}]{BPbook}%
  \BibitemOpen
  \bibfield  {author} {\bibinfo {author} {\bibfnamefont {H.-P.}\ \bibnamefont
  {Breuer}}\ and\ \bibinfo {author} {\bibfnamefont {F.}~\bibnamefont
  {Petruccione}},\ }\href@noop {} {\emph {\bibinfo {title} {The Theory of Open
  Quantum Systems}}}\ (\bibinfo  {publisher} {Oxford University Press},\
  \bibinfo {address} {Oxford},\ \bibinfo {year} {2007})\BibitemShut {NoStop}%
\bibitem [{\citenamefont {Benedetti}\ \emph {et~al.}(2014)\citenamefont
  {Benedetti}, \citenamefont {Buscemi}, \citenamefont {Bordone},\ and\
  \citenamefont {Paris}}]{ClaudiaPRA2014}%
  \BibitemOpen
  \bibfield  {author} {\bibinfo {author} {\bibfnamefont {C.}~\bibnamefont
  {Benedetti}}, \bibinfo {author} {\bibfnamefont {F.}~\bibnamefont {Buscemi}},
  \bibinfo {author} {\bibfnamefont {P.}~\bibnamefont {Bordone}},\ and\ \bibinfo
  {author} {\bibfnamefont {M.~G.~A.}\ \bibnamefont {Paris}},\ }\bibfield
  {title} {\bibinfo {title} {Quantum probes for the spectral properties of a
  classical environment},\ }\href {https://doi.org/10.1103/PhysRevA.89.032114}
  {\bibfield  {journal} {\bibinfo  {journal} {Phys. Rev. A}\ }\textbf {\bibinfo
  {volume} {89}},\ \bibinfo {pages} {032114} (\bibinfo {year}
  {2014})}\BibitemShut {NoStop}%
\bibitem [{\citenamefont {Elliott}\ and\ \citenamefont
  {Johnson}(2016)}]{Elliot2016}%
  \BibitemOpen
  \bibfield  {author} {\bibinfo {author} {\bibfnamefont {T.~J.}\ \bibnamefont
  {Elliott}}\ and\ \bibinfo {author} {\bibfnamefont {T.~H.}\ \bibnamefont
  {Johnson}},\ }\bibfield  {title} {\bibinfo {title} {Nondestructive probing of
  means, variances, and correlations of ultracold-atomic-system densities via
  qubit impurities},\ }\href {https://doi.org/10.1103/PhysRevA.93.043612}
  {\bibfield  {journal} {\bibinfo  {journal} {Phys. Rev. A}\ }\textbf {\bibinfo
  {volume} {93}},\ \bibinfo {pages} {043612} (\bibinfo {year}
  {2016})}\BibitemShut {NoStop}%
\bibitem [{\citenamefont {Norris}\ \emph {et~al.}(2016)\citenamefont {Norris},
  \citenamefont {Paz-Silva},\ and\ \citenamefont {Viola}}]{ViolaPRL2016}%
  \BibitemOpen
  \bibfield  {author} {\bibinfo {author} {\bibfnamefont {L.~M.}\ \bibnamefont
  {Norris}}, \bibinfo {author} {\bibfnamefont {G.~A.}\ \bibnamefont
  {Paz-Silva}},\ and\ \bibinfo {author} {\bibfnamefont {L.}~\bibnamefont
  {Viola}},\ }\bibfield  {title} {\bibinfo {title} {Qubit noise spectroscopy
  for non-gaussian dephasing environments},\ }\href
  {https://doi.org/10.1103/PhysRevLett.116.150503} {\bibfield  {journal}
  {\bibinfo  {journal} {Phys. Rev. Lett.}\ }\textbf {\bibinfo {volume} {116}},\
  \bibinfo {pages} {150503} (\bibinfo {year} {2016})}\BibitemShut {NoStop}%
\bibitem [{\citenamefont {Tamascelli}\ \emph {et~al.}(2016)\citenamefont
  {Tamascelli}, \citenamefont {Benedetti}, \citenamefont {Olivares},\ and\
  \citenamefont {Paris}}]{DarioPRA2016}%
  \BibitemOpen
  \bibfield  {author} {\bibinfo {author} {\bibfnamefont {D.}~\bibnamefont
  {Tamascelli}}, \bibinfo {author} {\bibfnamefont {C.}~\bibnamefont
  {Benedetti}}, \bibinfo {author} {\bibfnamefont {S.}~\bibnamefont
  {Olivares}},\ and\ \bibinfo {author} {\bibfnamefont {M.~G.~A.}\ \bibnamefont
  {Paris}},\ }\bibfield  {title} {\bibinfo {title} {Characterization of qubit
  chains by feynman probes},\ }\href
  {https://doi.org/10.1103/PhysRevA.94.042129} {\bibfield  {journal} {\bibinfo
  {journal} {Phys. Rev. A}\ }\textbf {\bibinfo {volume} {94}},\ \bibinfo
  {pages} {042129} (\bibinfo {year} {2016})}\BibitemShut {NoStop}%
\bibitem [{\citenamefont {Streif}\ \emph {et~al.}(2016)\citenamefont {Streif},
  \citenamefont {Buchleitner}, \citenamefont {Jaksch},\ and\ \citenamefont
  {Mur-Petit}}]{Streif2016}%
  \BibitemOpen
  \bibfield  {author} {\bibinfo {author} {\bibfnamefont {M.}~\bibnamefont
  {Streif}}, \bibinfo {author} {\bibfnamefont {A.}~\bibnamefont {Buchleitner}},
  \bibinfo {author} {\bibfnamefont {D.}~\bibnamefont {Jaksch}},\ and\ \bibinfo
  {author} {\bibfnamefont {J.}~\bibnamefont {Mur-Petit}},\ }\bibfield  {title}
  {\bibinfo {title} {Measuring correlations of cold-atom systems using multiple
  quantum probes},\ }\href {https://doi.org/10.1103/PhysRevA.94.053634}
  {\bibfield  {journal} {\bibinfo  {journal} {Phys. Rev. A}\ }\textbf {\bibinfo
  {volume} {94}},\ \bibinfo {pages} {053634} (\bibinfo {year}
  {2016})}\BibitemShut {NoStop}%
\bibitem [{\citenamefont {Benedetti}\ \emph {et~al.}(2018)\citenamefont
  {Benedetti}, \citenamefont {Salari~Sehdaran}, \citenamefont {Zandi},\ and\
  \citenamefont {Paris}}]{Benedetti2017}%
  \BibitemOpen
  \bibfield  {author} {\bibinfo {author} {\bibfnamefont {C.}~\bibnamefont
  {Benedetti}}, \bibinfo {author} {\bibfnamefont {F.}~\bibnamefont
  {Salari~Sehdaran}}, \bibinfo {author} {\bibfnamefont {M.~H.}\ \bibnamefont
  {Zandi}},\ and\ \bibinfo {author} {\bibfnamefont {M.~G.~A.}\ \bibnamefont
  {Paris}},\ }\bibfield  {title} {\bibinfo {title} {Quantum probes for the
  cutoff frequency of ohmic environments},\ }\href
  {https://doi.org/10.1103/PhysRevA.97.012126} {\bibfield  {journal} {\bibinfo
  {journal} {Phys. Rev. A}\ }\textbf {\bibinfo {volume} {97}},\ \bibinfo
  {pages} {012126} (\bibinfo {year} {2018})}\BibitemShut {NoStop}%
\bibitem [{\citenamefont {Cosco}\ \emph {et~al.}(2017)\citenamefont {Cosco},
  \citenamefont {Borrelli}, \citenamefont {Plastina},\ and\ \citenamefont
  {Maniscalco}}]{CoscoPRA2017}%
  \BibitemOpen
  \bibfield  {author} {\bibinfo {author} {\bibfnamefont {F.}~\bibnamefont
  {Cosco}}, \bibinfo {author} {\bibfnamefont {M.}~\bibnamefont {Borrelli}},
  \bibinfo {author} {\bibfnamefont {F.}~\bibnamefont {Plastina}},\ and\
  \bibinfo {author} {\bibfnamefont {S.}~\bibnamefont {Maniscalco}},\ }\bibfield
   {title} {\bibinfo {title} {Momentum-resolved and correlation spectroscopy
  using quantum probes},\ }\href {https://doi.org/10.1103/PhysRevA.95.053620}
  {\bibfield  {journal} {\bibinfo  {journal} {Phys. Rev. A}\ }\textbf {\bibinfo
  {volume} {95}},\ \bibinfo {pages} {053620} (\bibinfo {year}
  {2017})}\BibitemShut {NoStop}%
\bibitem [{\citenamefont {Sone}\ and\ \citenamefont
  {Cappellaro}(2017)}]{SonePRA2017}%
  \BibitemOpen
  \bibfield  {author} {\bibinfo {author} {\bibfnamefont {A.}~\bibnamefont
  {Sone}}\ and\ \bibinfo {author} {\bibfnamefont {P.}~\bibnamefont
  {Cappellaro}},\ }\bibfield  {title} {\bibinfo {title} {Exact dimension
  estimation of interacting qubit systems assisted by a single quantum probe},\
  }\href {https://doi.org/10.1103/PhysRevA.96.062334} {\bibfield  {journal}
  {\bibinfo  {journal} {Phys. Rev. A}\ }\textbf {\bibinfo {volume} {96}},\
  \bibinfo {pages} {062334} (\bibinfo {year} {2017})}\BibitemShut {NoStop}%
\bibitem [{\citenamefont {Gebbia}\ \emph {et~al.}(2020)\citenamefont {Gebbia},
  \citenamefont {Benedetti}, \citenamefont {Benatti}, \citenamefont
  {Floreanini}, \citenamefont {Bina},\ and\ \citenamefont
  {Paris}}]{ClaudiaPRA2020}%
  \BibitemOpen
  \bibfield  {author} {\bibinfo {author} {\bibfnamefont {F.}~\bibnamefont
  {Gebbia}}, \bibinfo {author} {\bibfnamefont {C.}~\bibnamefont {Benedetti}},
  \bibinfo {author} {\bibfnamefont {F.}~\bibnamefont {Benatti}}, \bibinfo
  {author} {\bibfnamefont {R.}~\bibnamefont {Floreanini}}, \bibinfo {author}
  {\bibfnamefont {M.}~\bibnamefont {Bina}},\ and\ \bibinfo {author}
  {\bibfnamefont {M.~G.~A.}\ \bibnamefont {Paris}},\ }\bibfield  {title}
  {\bibinfo {title} {Two-qubit quantum probes for the temperature of an ohmic
  environment},\ }\href {https://doi.org/10.1103/PhysRevA.101.032112}
  {\bibfield  {journal} {\bibinfo  {journal} {Phys. Rev. A}\ }\textbf {\bibinfo
  {volume} {101}},\ \bibinfo {pages} {032112} (\bibinfo {year}
  {2020})}\BibitemShut {NoStop}%
\bibitem [{\citenamefont {Wu}\ and\ \citenamefont {Shi}(2020)}]{WuPRA2020}%
  \BibitemOpen
  \bibfield  {author} {\bibinfo {author} {\bibfnamefont {W.}~\bibnamefont
  {Wu}}\ and\ \bibinfo {author} {\bibfnamefont {C.}~\bibnamefont {Shi}},\
  }\bibfield  {title} {\bibinfo {title} {Quantum parameter estimation in a
  dissipative environment},\ }\href
  {https://doi.org/10.1103/PhysRevA.102.032607} {\bibfield  {journal} {\bibinfo
   {journal} {Phys. Rev. A}\ }\textbf {\bibinfo {volume} {102}},\ \bibinfo
  {pages} {032607} (\bibinfo {year} {2020})}\BibitemShut {NoStop}%
\bibitem [{\citenamefont {Tamascelli}\ \emph {et~al.}(2020)\citenamefont
  {Tamascelli}, \citenamefont {Benedetti}, \citenamefont {Breuer},\ and\
  \citenamefont {Paris}}]{Tamascelli2020}%
  \BibitemOpen
  \bibfield  {author} {\bibinfo {author} {\bibfnamefont {D.}~\bibnamefont
  {Tamascelli}}, \bibinfo {author} {\bibfnamefont {C.}~\bibnamefont
  {Benedetti}}, \bibinfo {author} {\bibfnamefont {H.-P.}\ \bibnamefont
  {Breuer}},\ and\ \bibinfo {author} {\bibfnamefont {M.~G.~A.}\ \bibnamefont
  {Paris}},\ }\bibfield  {title} {\bibinfo {title} {Quantum probing beyond pure
  dephasing},\ }\href {https://doi.org/10.1088/1367-2630/aba0e5} {\bibfield
  {journal} {\bibinfo  {journal} {New J. Phys.}\ }\textbf {\bibinfo {volume}
  {22}},\ \bibinfo {pages} {083027} (\bibinfo {year} {2020})}\BibitemShut
  {NoStop}%
\bibitem [{\citenamefont {Gianani}\ \emph {et~al.}(2020)\citenamefont
  {Gianani}, \citenamefont {Farina}, \citenamefont {Barbieri}, \citenamefont
  {Cimini}, \citenamefont {Cavina},\ and\ \citenamefont
  {Giovannetti}}]{GiovannettiPhysRevR2020}%
  \BibitemOpen
  \bibfield  {author} {\bibinfo {author} {\bibfnamefont {I.}~\bibnamefont
  {Gianani}}, \bibinfo {author} {\bibfnamefont {D.}~\bibnamefont {Farina}},
  \bibinfo {author} {\bibfnamefont {M.}~\bibnamefont {Barbieri}}, \bibinfo
  {author} {\bibfnamefont {V.}~\bibnamefont {Cimini}}, \bibinfo {author}
  {\bibfnamefont {V.}~\bibnamefont {Cavina}},\ and\ \bibinfo {author}
  {\bibfnamefont {V.}~\bibnamefont {Giovannetti}},\ }\bibfield  {title}
  {\bibinfo {title} {Discrimination of thermal baths by single-qubit probes},\
  }\href {https://doi.org/10.1103/PhysRevResearch.2.033497} {\bibfield
  {journal} {\bibinfo  {journal} {Phys. Rev. Research}\ }\textbf {\bibinfo
  {volume} {2}},\ \bibinfo {pages} {033497} (\bibinfo {year}
  {2020})}\BibitemShut {NoStop}%
\bibitem [{\citenamefont {Helstrom}(1976)}]{Helstrombook}%
  \BibitemOpen
  \bibfield  {author} {\bibinfo {author} {\bibfnamefont {C.~W.}\ \bibnamefont
  {Helstrom}},\ }\href@noop {} {\emph {\bibinfo {title} {Quantum detection and
  Estimation Theory}}}\ (\bibinfo  {publisher} {Academic Press},\ \bibinfo
  {address} {New York},\ \bibinfo {year} {1976})\BibitemShut {NoStop}%
\bibitem [{\citenamefont {Fujiwara}(2001)}]{Fujiwara2001}%
  \BibitemOpen
  \bibfield  {author} {\bibinfo {author} {\bibfnamefont {A.}~\bibnamefont
  {Fujiwara}},\ }\bibfield  {title} {\bibinfo {title} {Quantum channel
  identification problem},\ }\href {https://doi.org/10.1103/PhysRevA.63.042304}
  {\bibfield  {journal} {\bibinfo  {journal} {Phys. Rev. A}\ }\textbf {\bibinfo
  {volume} {63}},\ \bibinfo {pages} {042304} (\bibinfo {year}
  {2001})}\BibitemShut {NoStop}%
\bibitem [{\citenamefont {Monras}(2006)}]{Monras2006}%
  \BibitemOpen
  \bibfield  {author} {\bibinfo {author} {\bibfnamefont {A.}~\bibnamefont
  {Monras}},\ }\bibfield  {title} {\bibinfo {title} {Optimal phase measurements
  with pure gaussian states},\ }\href
  {https://doi.org/10.1103/PhysRevA.73.033821} {\bibfield  {journal} {\bibinfo
  {journal} {Phys. Rev. A}\ }\textbf {\bibinfo {volume} {73}},\ \bibinfo
  {pages} {033821} (\bibinfo {year} {2006})}\BibitemShut {NoStop}%
\bibitem [{\citenamefont {Monras}\ and\ \citenamefont
  {Paris}(2007)}]{Monras2007}%
  \BibitemOpen
  \bibfield  {author} {\bibinfo {author} {\bibfnamefont {A.}~\bibnamefont
  {Monras}}\ and\ \bibinfo {author} {\bibfnamefont {M.~G.~A.}\ \bibnamefont
  {Paris}},\ }\bibfield  {title} {\bibinfo {title} {Optimal quantum estimation
  of loss in bosonic channels},\ }\href
  {https://doi.org/10.1103/PhysRevLett.98.160401} {\bibfield  {journal}
  {\bibinfo  {journal} {Phys. Rev. Lett.}\ }\textbf {\bibinfo {volume} {98}},\
  \bibinfo {pages} {160401} (\bibinfo {year} {2007})}\BibitemShut {NoStop}%
\bibitem [{\citenamefont {Paris}(2008)}]{Paris2008QuantumEF}%
  \BibitemOpen
  \bibfield  {author} {\bibinfo {author} {\bibfnamefont {M.~G.~A.}\
  \bibnamefont {Paris}},\ }\bibfield  {title} {\bibinfo {title} {Quantum
  estimation for quantum technology},\ }\href@noop {} {\bibfield  {journal}
  {\bibinfo  {journal} {International Journal of Quantum Information}\ }\textbf
  {\bibinfo {volume} {07}},\ \bibinfo {pages} {125} (\bibinfo {year}
  {2008})}\BibitemShut {NoStop}%
\bibitem [{\citenamefont {Genoni}\ \emph {et~al.}(2011)\citenamefont {Genoni},
  \citenamefont {Olivares},\ and\ \citenamefont {Paris}}]{Genoni2011}%
  \BibitemOpen
  \bibfield  {author} {\bibinfo {author} {\bibfnamefont {M.~G.}\ \bibnamefont
  {Genoni}}, \bibinfo {author} {\bibfnamefont {S.}~\bibnamefont {Olivares}},\
  and\ \bibinfo {author} {\bibfnamefont {M.~G.~A.}\ \bibnamefont {Paris}},\
  }\bibfield  {title} {\bibinfo {title} {Optical phase estimation in the
  presence of phase diffusion},\ }\href
  {https://doi.org/10.1103/PhysRevLett.106.153603} {\bibfield  {journal}
  {\bibinfo  {journal} {Phys. Rev. Lett.}\ }\textbf {\bibinfo {volume} {106}},\
  \bibinfo {pages} {153603} (\bibinfo {year} {2011})}\BibitemShut {NoStop}%
\bibitem [{\citenamefont {Spagnolo}\ \emph {et~al.}(2012)\citenamefont
  {Spagnolo}, \citenamefont {Vitelli}, \citenamefont {Lucivero}, \citenamefont
  {Giovannetti}, \citenamefont {Maccone},\ and\ \citenamefont
  {Sciarrino}}]{Spagnolo2012}%
  \BibitemOpen
  \bibfield  {author} {\bibinfo {author} {\bibfnamefont {N.}~\bibnamefont
  {Spagnolo}}, \bibinfo {author} {\bibfnamefont {C.}~\bibnamefont {Vitelli}},
  \bibinfo {author} {\bibfnamefont {V.~G.}\ \bibnamefont {Lucivero}}, \bibinfo
  {author} {\bibfnamefont {V.}~\bibnamefont {Giovannetti}}, \bibinfo {author}
  {\bibfnamefont {L.}~\bibnamefont {Maccone}},\ and\ \bibinfo {author}
  {\bibfnamefont {F.}~\bibnamefont {Sciarrino}},\ }\bibfield  {title} {\bibinfo
  {title} {Phase estimation via quantum interferometry for noisy detectors},\
  }\href {https://doi.org/10.1103/PhysRevLett.108.233602} {\bibfield  {journal}
  {\bibinfo  {journal} {Phys. Rev. Lett.}\ }\textbf {\bibinfo {volume} {108}},\
  \bibinfo {pages} {233602} (\bibinfo {year} {2012})}\BibitemShut {NoStop}%
\bibitem [{\citenamefont {Pinel}\ \emph {et~al.}(2013)\citenamefont {Pinel},
  \citenamefont {Jian}, \citenamefont {Treps}, \citenamefont {Fabre},\ and\
  \citenamefont {Braun}}]{Pinel2013}%
  \BibitemOpen
  \bibfield  {author} {\bibinfo {author} {\bibfnamefont {O.}~\bibnamefont
  {Pinel}}, \bibinfo {author} {\bibfnamefont {P.}~\bibnamefont {Jian}},
  \bibinfo {author} {\bibfnamefont {N.}~\bibnamefont {Treps}}, \bibinfo
  {author} {\bibfnamefont {C.}~\bibnamefont {Fabre}},\ and\ \bibinfo {author}
  {\bibfnamefont {D.}~\bibnamefont {Braun}},\ }\bibfield  {title} {\bibinfo
  {title} {Quantum parameter estimation using general single-mode gaussian
  states},\ }\href {https://doi.org/10.1103/PhysRevA.88.040102} {\bibfield
  {journal} {\bibinfo  {journal} {Phys. Rev. A}\ }\textbf {\bibinfo {volume}
  {88}},\ \bibinfo {pages} {040102(R)} (\bibinfo {year} {2013})}\BibitemShut
  {NoStop}%
\bibitem [{\citenamefont {Chaudhry}(2014)}]{ChaudhryNV2014}%
  \BibitemOpen
  \bibfield  {author} {\bibinfo {author} {\bibfnamefont {A.~Z.}\ \bibnamefont
  {Chaudhry}},\ }\bibfield  {title} {\bibinfo {title} {Utilizing
  nitrogen-vacancy centers to measure oscillating magnetic fields},\ }\href
  {https://doi.org/10.1103/PhysRevA.90.042104} {\bibfield  {journal} {\bibinfo
  {journal} {Phys. Rev. A}\ }\textbf {\bibinfo {volume} {90}},\ \bibinfo
  {pages} {042104} (\bibinfo {year} {2014})}\BibitemShut {NoStop}%
\bibitem [{\citenamefont {Chaudhry}(2015)}]{ChaudhryNV2015}%
  \BibitemOpen
  \bibfield  {author} {\bibinfo {author} {\bibfnamefont {A.~Z.}\ \bibnamefont
  {Chaudhry}},\ }\bibfield  {title} {\bibinfo {title} {Detecting the presence
  of weak magnetic fields using nitrogen-vacancy centers},\ }\href
  {https://doi.org/10.1103/PhysRevA.91.062111} {\bibfield  {journal} {\bibinfo
  {journal} {Phys. Rev. A}\ }\textbf {\bibinfo {volume} {91}},\ \bibinfo
  {pages} {062111} (\bibinfo {year} {2015})}\BibitemShut {NoStop}%
\bibitem [{\citenamefont {Hakim}\ and\ \citenamefont
  {Ambegaokar}(1985)}]{HakimPRA1985}%
  \BibitemOpen
  \bibfield  {author} {\bibinfo {author} {\bibfnamefont {V.}~\bibnamefont
  {Hakim}}\ and\ \bibinfo {author} {\bibfnamefont {V.}~\bibnamefont
  {Ambegaokar}},\ }\bibfield  {title} {\bibinfo {title} {Quantum theory of a
  free particle interacting with a linearly dissipative environment},\ }\href
  {https://doi.org/10.1103/PhysRevA.32.423} {\bibfield  {journal} {\bibinfo
  {journal} {Phys. Rev. A}\ }\textbf {\bibinfo {volume} {32}},\ \bibinfo
  {pages} {423} (\bibinfo {year} {1985})}\BibitemShut {NoStop}%
\bibitem [{\citenamefont {Haake}\ and\ \citenamefont
  {Reibold}(1985)}]{HaakePRA1985}%
  \BibitemOpen
  \bibfield  {author} {\bibinfo {author} {\bibfnamefont {F.}~\bibnamefont
  {Haake}}\ and\ \bibinfo {author} {\bibfnamefont {R.}~\bibnamefont
  {Reibold}},\ }\bibfield  {title} {\bibinfo {title} {Strong damping and
  low-temperature anomalies for the harmonic oscillator},\ }\href
  {https://doi.org/10.1103/PhysRevA.32.2462} {\bibfield  {journal} {\bibinfo
  {journal} {Phys. Rev. A}\ }\textbf {\bibinfo {volume} {32}},\ \bibinfo
  {pages} {2462} (\bibinfo {year} {1985})}\BibitemShut {NoStop}%
\bibitem [{\citenamefont {Grabert}\ \emph {et~al.}(1988)\citenamefont
  {Grabert}, \citenamefont {Schramm},\ and\ \citenamefont
  {Ingold}}]{Grabert1988}%
  \BibitemOpen
  \bibfield  {author} {\bibinfo {author} {\bibfnamefont {H.}~\bibnamefont
  {Grabert}}, \bibinfo {author} {\bibfnamefont {P.}~\bibnamefont {Schramm}},\
  and\ \bibinfo {author} {\bibfnamefont {G.-L.}\ \bibnamefont {Ingold}},\
  }\bibfield  {title} {\bibinfo {title} {Quantum brownian motion: The
  functional integral approach},\ }\href
  {https://doi.org/10.1016/0370-1573(88)90023-3} {\bibfield  {journal}
  {\bibinfo  {journal} {Phys. Rep.}\ }\textbf {\bibinfo {volume} {168}},\
  \bibinfo {pages} {115} (\bibinfo {year} {1988})}\BibitemShut {NoStop}%
\bibitem [{\citenamefont {Smith}\ and\ \citenamefont
  {Caldeira}(1990)}]{SmithPRA1990}%
  \BibitemOpen
  \bibfield  {author} {\bibinfo {author} {\bibfnamefont {C.~M.}\ \bibnamefont
  {Smith}}\ and\ \bibinfo {author} {\bibfnamefont {A.~O.}\ \bibnamefont
  {Caldeira}},\ }\bibfield  {title} {\bibinfo {title} {Application of the
  generalized feynman-vernon approach to a simple system: The damped harmonic
  oscillator},\ }\href {https://doi.org/10.1103/PhysRevA.41.3103} {\bibfield
  {journal} {\bibinfo  {journal} {Phys. Rev. A}\ }\textbf {\bibinfo {volume}
  {41}},\ \bibinfo {pages} {3103} (\bibinfo {year} {1990})}\BibitemShut
  {NoStop}%
\bibitem [{\citenamefont {Karrlein}\ and\ \citenamefont
  {Grabert}(1997)}]{GrabertPRE1997}%
  \BibitemOpen
  \bibfield  {author} {\bibinfo {author} {\bibfnamefont {R.}~\bibnamefont
  {Karrlein}}\ and\ \bibinfo {author} {\bibfnamefont {H.}~\bibnamefont
  {Grabert}},\ }\bibfield  {title} {\bibinfo {title} {Exact time evolution and
  master equations for the damped harmonic oscillator},\ }\href
  {https://doi.org/10.1103/PhysRevE.55.153} {\bibfield  {journal} {\bibinfo
  {journal} {Phys. Rev. E}\ }\textbf {\bibinfo {volume} {55}},\ \bibinfo
  {pages} {153} (\bibinfo {year} {1997})}\BibitemShut {NoStop}%
\bibitem [{\citenamefont {Romero}\ and\ \citenamefont
  {Paz}(1997)}]{PazPRA1997}%
  \BibitemOpen
  \bibfield  {author} {\bibinfo {author} {\bibfnamefont {L.~D.}\ \bibnamefont
  {Romero}}\ and\ \bibinfo {author} {\bibfnamefont {J.~P.}\ \bibnamefont
  {Paz}},\ }\bibfield  {title} {\bibinfo {title} {Decoherence and initial
  correlations in quantum brownian motion},\ }\href
  {https://doi.org/10.1103/PhysRevA.55.4070} {\bibfield  {journal} {\bibinfo
  {journal} {Phys. Rev. A}\ }\textbf {\bibinfo {volume} {55}},\ \bibinfo
  {pages} {4070} (\bibinfo {year} {1997})}\BibitemShut {NoStop}%
\bibitem [{\citenamefont {Lutz}(2003)}]{LutzPRA2003}%
  \BibitemOpen
  \bibfield  {author} {\bibinfo {author} {\bibfnamefont {E.}~\bibnamefont
  {Lutz}},\ }\bibfield  {title} {\bibinfo {title} {Effect of initial
  correlations on short-time decoherence},\ }\href
  {https://doi.org/10.1103/PhysRevA.67.022109} {\bibfield  {journal} {\bibinfo
  {journal} {Phys. Rev. A}\ }\textbf {\bibinfo {volume} {67}},\ \bibinfo
  {pages} {022109} (\bibinfo {year} {2003})}\BibitemShut {NoStop}%
\bibitem [{\citenamefont {Banerjee}\ and\ \citenamefont
  {Ghosh}(2003)}]{BanerjeePRE2003}%
  \BibitemOpen
  \bibfield  {author} {\bibinfo {author} {\bibfnamefont {S.}~\bibnamefont
  {Banerjee}}\ and\ \bibinfo {author} {\bibfnamefont {R.}~\bibnamefont
  {Ghosh}},\ }\bibfield  {title} {\bibinfo {title} {General quantum brownian
  motion with initially correlated and nonlinearly coupled environment},\
  }\href {https://doi.org/10.1103/PhysRevE.67.056120} {\bibfield  {journal}
  {\bibinfo  {journal} {Phys. Rev. E}\ }\textbf {\bibinfo {volume} {67}},\
  \bibinfo {pages} {056120} (\bibinfo {year} {2003})}\BibitemShut {NoStop}%
\bibitem [{\citenamefont {van Kampen}(2004)}]{vanKampen2004}%
  \BibitemOpen
  \bibfield  {author} {\bibinfo {author} {\bibfnamefont {N.~G.}\ \bibnamefont
  {van Kampen}},\ }\href {https://doi.org/10.1023/B:JOSS.0000022383.06086.6c}
  {\bibfield  {journal} {\bibinfo  {journal} {J. Stat. Phys.}\ }\textbf
  {\bibinfo {volume} {115}},\ \bibinfo {pages} {1057} (\bibinfo {year}
  {2004})}\BibitemShut {NoStop}%
\bibitem [{\citenamefont {Ban}(2009)}]{BanPRA2009}%
  \BibitemOpen
  \bibfield  {author} {\bibinfo {author} {\bibfnamefont {M.}~\bibnamefont
  {Ban}},\ }\bibfield  {title} {\bibinfo {title} {Quantum master equation for
  dephasing of a two-level system with an initial correlation},\ }\href
  {https://doi.org/10.1103/PhysRevA.80.064103} {\bibfield  {journal} {\bibinfo
  {journal} {Phys. Rev. A}\ }\textbf {\bibinfo {volume} {80}},\ \bibinfo
  {pages} {064103} (\bibinfo {year} {2009})}\BibitemShut {NoStop}%
\bibitem [{\citenamefont {Campisi}\ \emph {et~al.}(2009)\citenamefont
  {Campisi}, \citenamefont {Talkner},\ and\ \citenamefont
  {H\"anggi}}]{HanggiPRL2009}%
  \BibitemOpen
  \bibfield  {author} {\bibinfo {author} {\bibfnamefont {M.}~\bibnamefont
  {Campisi}}, \bibinfo {author} {\bibfnamefont {P.}~\bibnamefont {Talkner}},\
  and\ \bibinfo {author} {\bibfnamefont {P.}~\bibnamefont {H\"anggi}},\
  }\bibfield  {title} {\bibinfo {title} {Fluctuation theorem for arbitrary open
  quantum systems},\ }\href {https://doi.org/10.1103/PhysRevLett.102.210401}
  {\bibfield  {journal} {\bibinfo  {journal} {Phys. Rev. Lett.}\ }\textbf
  {\bibinfo {volume} {102}},\ \bibinfo {pages} {210401} (\bibinfo {year}
  {2009})}\BibitemShut {NoStop}%
\bibitem [{\citenamefont {Uchiyama}\ and\ \citenamefont
  {Aihara}(2010)}]{UchiyamaPRA2010}%
  \BibitemOpen
  \bibfield  {author} {\bibinfo {author} {\bibfnamefont {C.}~\bibnamefont
  {Uchiyama}}\ and\ \bibinfo {author} {\bibfnamefont {M.}~\bibnamefont
  {Aihara}},\ }\bibfield  {title} {\bibinfo {title} {Role of initial quantum
  correlation in transient linear response},\ }\href
  {https://doi.org/10.1103/PhysRevA.82.044104} {\bibfield  {journal} {\bibinfo
  {journal} {Phys. Rev. A}\ }\textbf {\bibinfo {volume} {82}},\ \bibinfo
  {pages} {044104} (\bibinfo {year} {2010})}\BibitemShut {NoStop}%
\bibitem [{\citenamefont {Dijkstra}\ and\ \citenamefont
  {Tanimura}(2010)}]{TanimuraPRL2010}%
  \BibitemOpen
  \bibfield  {author} {\bibinfo {author} {\bibfnamefont {A.~G.}\ \bibnamefont
  {Dijkstra}}\ and\ \bibinfo {author} {\bibfnamefont {Y.}~\bibnamefont
  {Tanimura}},\ }\bibfield  {title} {\bibinfo {title} {Non-markovian
  entanglement dynamics in the presence of system-bath coherence},\ }\href
  {https://doi.org/10.1103/PhysRevLett.104.250401} {\bibfield  {journal}
  {\bibinfo  {journal} {Phys. Rev. Lett.}\ }\textbf {\bibinfo {volume} {104}},\
  \bibinfo {pages} {250401} (\bibinfo {year} {2010})}\BibitemShut {NoStop}%
\bibitem [{\citenamefont {Smirne}\ \emph {et~al.}(2010)\citenamefont {Smirne},
  \citenamefont {Breuer}, \citenamefont {Piilo},\ and\ \citenamefont
  {Vacchini}}]{SmirnePRA2010}%
  \BibitemOpen
  \bibfield  {author} {\bibinfo {author} {\bibfnamefont {A.}~\bibnamefont
  {Smirne}}, \bibinfo {author} {\bibfnamefont {H.-P.}\ \bibnamefont {Breuer}},
  \bibinfo {author} {\bibfnamefont {J.}~\bibnamefont {Piilo}},\ and\ \bibinfo
  {author} {\bibfnamefont {B.}~\bibnamefont {Vacchini}},\ }\bibfield  {title}
  {\bibinfo {title} {Initial correlations in open-systems dynamics: The
  jaynes-cummings model},\ }\href {https://doi.org/10.1103/PhysRevA.82.062114}
  {\bibfield  {journal} {\bibinfo  {journal} {Phys. Rev. A}\ }\textbf {\bibinfo
  {volume} {82}},\ \bibinfo {pages} {062114} (\bibinfo {year}
  {2010})}\BibitemShut {NoStop}%
\bibitem [{\citenamefont {Dajka}\ and\ \citenamefont
  {\L{}uczka}(2010)}]{DajkaPRA2010}%
  \BibitemOpen
  \bibfield  {author} {\bibinfo {author} {\bibfnamefont {J.}~\bibnamefont
  {Dajka}}\ and\ \bibinfo {author} {\bibfnamefont {J.}~\bibnamefont
  {\L{}uczka}},\ }\bibfield  {title} {\bibinfo {title} {Distance growth of
  quantum states due to initial system-environment correlations},\ }\href
  {https://doi.org/10.1103/PhysRevA.82.012341} {\bibfield  {journal} {\bibinfo
  {journal} {Phys. Rev. A}\ }\textbf {\bibinfo {volume} {82}},\ \bibinfo
  {pages} {012341} (\bibinfo {year} {2010})}\BibitemShut {NoStop}%
\bibitem [{\citenamefont {Zhang}\ \emph {et~al.}(2010)\citenamefont {Zhang},
  \citenamefont {Zou}, \citenamefont {Xia},\ and\ \citenamefont
  {Guo}}]{ZhangPRA2010}%
  \BibitemOpen
  \bibfield  {author} {\bibinfo {author} {\bibfnamefont {Y.-J.}\ \bibnamefont
  {Zhang}}, \bibinfo {author} {\bibfnamefont {X.-B.}\ \bibnamefont {Zou}},
  \bibinfo {author} {\bibfnamefont {Y.-J.}\ \bibnamefont {Xia}},\ and\ \bibinfo
  {author} {\bibfnamefont {G.-C.}\ \bibnamefont {Guo}},\ }\bibfield  {title}
  {\bibinfo {title} {Different entanglement dynamical behaviors due to initial
  system-environment correlations},\ }\href
  {https://doi.org/10.1103/PhysRevA.82.022108} {\bibfield  {journal} {\bibinfo
  {journal} {Phys. Rev. A}\ }\textbf {\bibinfo {volume} {82}},\ \bibinfo
  {pages} {022108} (\bibinfo {year} {2010})}\BibitemShut {NoStop}%
\bibitem [{\citenamefont {Tan}\ and\ \citenamefont {Zhang}(2011)}]{TanPRA2011}%
  \BibitemOpen
  \bibfield  {author} {\bibinfo {author} {\bibfnamefont {H.-T.}\ \bibnamefont
  {Tan}}\ and\ \bibinfo {author} {\bibfnamefont {W.-M.}\ \bibnamefont
  {Zhang}},\ }\bibfield  {title} {\bibinfo {title} {Non-markovian dynamics of
  an open quantum system with initial system-reservoir correlations: A
  nanocavity coupled to a coupled-resonator optical waveguide},\ }\href
  {https://doi.org/10.1103/PhysRevA.83.032102} {\bibfield  {journal} {\bibinfo
  {journal} {Phys. Rev. A}\ }\textbf {\bibinfo {volume} {83}},\ \bibinfo
  {pages} {032102} (\bibinfo {year} {2011})}\BibitemShut {NoStop}%
\bibitem [{\citenamefont {Lee}\ \emph {et~al.}(2012)\citenamefont {Lee},
  \citenamefont {Cao},\ and\ \citenamefont {Gong}}]{CKLeePRE2012}%
  \BibitemOpen
  \bibfield  {author} {\bibinfo {author} {\bibfnamefont {C.~K.}\ \bibnamefont
  {Lee}}, \bibinfo {author} {\bibfnamefont {J.}~\bibnamefont {Cao}},\ and\
  \bibinfo {author} {\bibfnamefont {J.}~\bibnamefont {Gong}},\ }\bibfield
  {title} {\bibinfo {title} {Noncanonical statistics of a spin-boson model:
  Theory and exact monte carlo simulations},\ }\href
  {https://doi.org/10.1103/PhysRevE.86.021109} {\bibfield  {journal} {\bibinfo
  {journal} {Phys. Rev. E}\ }\textbf {\bibinfo {volume} {86}},\ \bibinfo
  {pages} {021109} (\bibinfo {year} {2012})}\BibitemShut {NoStop}%
\bibitem [{\citenamefont {Morozov}\ \emph {et~al.}(2012)\citenamefont
  {Morozov}, \citenamefont {Mathey},\ and\ \citenamefont
  {R\"opke}}]{MorozovPRA2012}%
  \BibitemOpen
  \bibfield  {author} {\bibinfo {author} {\bibfnamefont {V.~G.}\ \bibnamefont
  {Morozov}}, \bibinfo {author} {\bibfnamefont {S.}~\bibnamefont {Mathey}},\
  and\ \bibinfo {author} {\bibfnamefont {G.}~\bibnamefont {R\"opke}},\
  }\bibfield  {title} {\bibinfo {title} {Decoherence in an exactly solvable
  qubit model with initial qubit-environment correlations},\ }\href
  {https://doi.org/10.1103/PhysRevA.85.022101} {\bibfield  {journal} {\bibinfo
  {journal} {Phys. Rev. A}\ }\textbf {\bibinfo {volume} {85}},\ \bibinfo
  {pages} {022101} (\bibinfo {year} {2012})}\BibitemShut {NoStop}%
\bibitem [{\citenamefont {Semin}\ \emph {et~al.}(2012)\citenamefont {Semin},
  \citenamefont {Sinayskiy},\ and\ \citenamefont {Petruccione}}]{SeminPRA2012}%
  \BibitemOpen
  \bibfield  {author} {\bibinfo {author} {\bibfnamefont {V.}~\bibnamefont
  {Semin}}, \bibinfo {author} {\bibfnamefont {I.}~\bibnamefont {Sinayskiy}},\
  and\ \bibinfo {author} {\bibfnamefont {F.}~\bibnamefont {Petruccione}},\
  }\bibfield  {title} {\bibinfo {title} {Initial correlation in a system of a
  spin coupled to a spin bath through an intermediate spin},\ }\href
  {https://doi.org/10.1103/PhysRevA.86.062114} {\bibfield  {journal} {\bibinfo
  {journal} {Phys. Rev. A}\ }\textbf {\bibinfo {volume} {86}},\ \bibinfo
  {pages} {062114} (\bibinfo {year} {2012})}\BibitemShut {NoStop}%
\bibitem [{\citenamefont {Chaudhry}\ and\ \citenamefont
  {Gong}(2013{\natexlab{a}})}]{ChaudhryPRA2013a}%
  \BibitemOpen
  \bibfield  {author} {\bibinfo {author} {\bibfnamefont {A.~Z.}\ \bibnamefont
  {Chaudhry}}\ and\ \bibinfo {author} {\bibfnamefont {J.}~\bibnamefont
  {Gong}},\ }\bibfield  {title} {\bibinfo {title} {Amplification and
  suppression of system-bath-correlation effects in an open many-body system},\
  }\href {https://doi.org/10.1103/PhysRevA.87.012129} {\bibfield  {journal}
  {\bibinfo  {journal} {Phys. Rev. A}\ }\textbf {\bibinfo {volume} {87}},\
  \bibinfo {pages} {012129} (\bibinfo {year} {2013}{\natexlab{a}})}\BibitemShut
  {NoStop}%
\bibitem [{\citenamefont {Chaudhry}\ and\ \citenamefont
  {Gong}(2013{\natexlab{b}})}]{ChaudhryPRA2013b}%
  \BibitemOpen
  \bibfield  {author} {\bibinfo {author} {\bibfnamefont {A.~Z.}\ \bibnamefont
  {Chaudhry}}\ and\ \bibinfo {author} {\bibfnamefont {J.}~\bibnamefont
  {Gong}},\ }\bibfield  {title} {\bibinfo {title} {Role of initial
  system-environment correlations: A master equation approach},\ }\href
  {https://doi.org/10.1103/PhysRevA.88.052107} {\bibfield  {journal} {\bibinfo
  {journal} {Phys. Rev. A}\ }\textbf {\bibinfo {volume} {88}},\ \bibinfo
  {pages} {052107} (\bibinfo {year} {2013}{\natexlab{b}})}\BibitemShut
  {NoStop}%
\bibitem [{\citenamefont {Chaudhry}\ and\ \citenamefont
  {Gong}(2013{\natexlab{c}})}]{ChaudhryCJC2013}%
  \BibitemOpen
  \bibfield  {author} {\bibinfo {author} {\bibfnamefont {A.~Z.}\ \bibnamefont
  {Chaudhry}}\ and\ \bibinfo {author} {\bibfnamefont {J.}~\bibnamefont
  {Gong}},\ }\bibfield  {title} {\bibinfo {title} {The effect of state
  preparation in a many-body system},\ }\href@noop {} {\bibfield  {journal}
  {\bibinfo  {journal} {Can. J. Chem.}\ }\textbf {\bibinfo {volume} {92}},\
  \bibinfo {pages} {119} (\bibinfo {year} {2013}{\natexlab{c}})}\BibitemShut
  {NoStop}%
\bibitem [{\citenamefont {Reina}\ \emph {et~al.}(2014)\citenamefont {Reina},
  \citenamefont {Susa},\ and\ \citenamefont {Fanchini}}]{FanchiniSciRep2014}%
  \BibitemOpen
  \bibfield  {author} {\bibinfo {author} {\bibfnamefont {J.}~\bibnamefont
  {Reina}}, \bibinfo {author} {\bibfnamefont {C.}~\bibnamefont {Susa}},\ and\
  \bibinfo {author} {\bibfnamefont {F.}~\bibnamefont {Fanchini}},\ }\bibfield
  {title} {\bibinfo {title} {Extracting information from qubit-environment
  correlations},\ }\href {https://doi.org/10.1038/srep07443} {\bibfield
  {journal} {\bibinfo  {journal} {Sci. Rep.}\ }\textbf {\bibinfo {volume}
  {4}},\ \bibinfo {pages} {7443} (\bibinfo {year} {2014})}\BibitemShut
  {NoStop}%
\bibitem [{\citenamefont {Zhang}\ \emph {et~al.}(2015)\citenamefont {Zhang},
  \citenamefont {Han}, \citenamefont {Xia}, \citenamefont {Yu},\ and\
  \citenamefont {Fan}}]{FanSciRep2015correlation}%
  \BibitemOpen
  \bibfield  {author} {\bibinfo {author} {\bibfnamefont {Y.-J.}\ \bibnamefont
  {Zhang}}, \bibinfo {author} {\bibfnamefont {W.}~\bibnamefont {Han}}, \bibinfo
  {author} {\bibfnamefont {Y.-J.}\ \bibnamefont {Xia}}, \bibinfo {author}
  {\bibfnamefont {Y.-M.}\ \bibnamefont {Yu}},\ and\ \bibinfo {author}
  {\bibfnamefont {H.}~\bibnamefont {Fan}},\ }\bibfield  {title} {\bibinfo
  {title} {Role of initial system-bath correlation on coherence trapping},\
  }\href {https://doi.org/10.1038/srep13359} {\bibfield  {journal} {\bibinfo
  {journal} {Sci. Rep.}\ }\textbf {\bibinfo {volume} {5}},\ \bibinfo {pages}
  {13359} (\bibinfo {year} {2015})}\BibitemShut {NoStop}%
\bibitem [{\citenamefont {Chen}\ and\ \citenamefont
  {Goan}(2016)}]{ChenPRA2016}%
  \BibitemOpen
  \bibfield  {author} {\bibinfo {author} {\bibfnamefont {C.-C.}\ \bibnamefont
  {Chen}}\ and\ \bibinfo {author} {\bibfnamefont {H.-S.}\ \bibnamefont
  {Goan}},\ }\bibfield  {title} {\bibinfo {title} {Effects of initial
  system-environment correlations on open-quantum-system dynamics and state
  preparation},\ }\href {https://doi.org/10.1103/PhysRevA.93.032113} {\bibfield
   {journal} {\bibinfo  {journal} {Phys. Rev. A}\ }\textbf {\bibinfo {volume}
  {93}},\ \bibinfo {pages} {032113} (\bibinfo {year} {2016})}\BibitemShut
  {NoStop}%
\bibitem [{\citenamefont {de~Vega}\ and\ \citenamefont
  {Alonso}(2017)}]{VegaRMP2017}%
  \BibitemOpen
  \bibfield  {author} {\bibinfo {author} {\bibfnamefont {I.}~\bibnamefont
  {de~Vega}}\ and\ \bibinfo {author} {\bibfnamefont {D.}~\bibnamefont
  {Alonso}},\ }\bibfield  {title} {\bibinfo {title} {Dynamics of non-markovian
  open quantum systems},\ }\href {https://doi.org/10.1103/RevModPhys.89.015001}
  {\bibfield  {journal} {\bibinfo  {journal} {Rev. Mod. Phys.}\ }\textbf
  {\bibinfo {volume} {89}},\ \bibinfo {pages} {015001} (\bibinfo {year}
  {2017})}\BibitemShut {NoStop}%
\bibitem [{\citenamefont {Halimeh}\ and\ \citenamefont
  {de~Vega}(2017)}]{VegaPRA2017}%
  \BibitemOpen
  \bibfield  {author} {\bibinfo {author} {\bibfnamefont {J.~C.}\ \bibnamefont
  {Halimeh}}\ and\ \bibinfo {author} {\bibfnamefont {I.}~\bibnamefont
  {de~Vega}},\ }\bibfield  {title} {\bibinfo {title} {Weak-coupling master
  equation for arbitrary initial conditions},\ }\href
  {https://doi.org/10.1103/PhysRevA.95.052108} {\bibfield  {journal} {\bibinfo
  {journal} {Phys. Rev. A}\ }\textbf {\bibinfo {volume} {95}},\ \bibinfo
  {pages} {052108} (\bibinfo {year} {2017})}\BibitemShut {NoStop}%
\bibitem [{\citenamefont {Kitajima}\ \emph {et~al.}(2017)\citenamefont
  {Kitajima}, \citenamefont {Ban},\ and\ \citenamefont
  {Shibata}}]{ShibataJPhysA2017}%
  \BibitemOpen
  \bibfield  {author} {\bibinfo {author} {\bibfnamefont {S.}~\bibnamefont
  {Kitajima}}, \bibinfo {author} {\bibfnamefont {M.}~\bibnamefont {Ban}},\ and\
  \bibinfo {author} {\bibfnamefont {F.}~\bibnamefont {Shibata}},\ }\bibfield
  {title} {\bibinfo {title} {Expansion formulas for quantum master equations
  including initial correlation},\ }\href@noop {} {\bibfield  {journal}
  {\bibinfo  {journal} {J. Phys. A: Math. Theor}\ }\textbf {\bibinfo {volume}
  {50}},\ \bibinfo {pages} {125303} (\bibinfo {year} {2017})}\BibitemShut
  {NoStop}%
\bibitem [{\citenamefont {Buser}\ \emph {et~al.}(2017)\citenamefont {Buser},
  \citenamefont {Cerrillo}, \citenamefont {Schaller},\ and\ \citenamefont
  {Cao}}]{CaoPRA2017}%
  \BibitemOpen
  \bibfield  {author} {\bibinfo {author} {\bibfnamefont {M.}~\bibnamefont
  {Buser}}, \bibinfo {author} {\bibfnamefont {J.}~\bibnamefont {Cerrillo}},
  \bibinfo {author} {\bibfnamefont {G.}~\bibnamefont {Schaller}},\ and\
  \bibinfo {author} {\bibfnamefont {J.}~\bibnamefont {Cao}},\ }\bibfield
  {title} {\bibinfo {title} {Initial system-environment correlations via the
  transfer-tensor method},\ }\href {https://doi.org/10.1103/PhysRevA.96.062122}
  {\bibfield  {journal} {\bibinfo  {journal} {Phys. Rev. A}\ }\textbf {\bibinfo
  {volume} {96}},\ \bibinfo {pages} {062122} (\bibinfo {year}
  {2017})}\BibitemShut {NoStop}%
\bibitem [{\citenamefont {Majeed}\ and\ \citenamefont
  {Chaudhry}(2019)}]{MehwishEurPhysJD2019}%
  \BibitemOpen
  \bibfield  {author} {\bibinfo {author} {\bibfnamefont {M.}~\bibnamefont
  {Majeed}}\ and\ \bibinfo {author} {\bibfnamefont {A.~Z.}\ \bibnamefont
  {Chaudhry}},\ }\bibfield  {title} {\bibinfo {title} {Effect of initial
  system–environment correlations with spin environments},\ }\href@noop {}
  {\bibfield  {journal} {\bibinfo  {journal} {Eur. Phys. J. D}\ }\textbf
  {\bibinfo {volume} {73}},\ \bibinfo {pages} {16} (\bibinfo {year}
  {2019})}\BibitemShut {NoStop}%
\bibitem [{\citenamefont {Viola}\ and\ \citenamefont
  {Lloyd}(1998)}]{ViolaPRA1998}%
  \BibitemOpen
  \bibfield  {author} {\bibinfo {author} {\bibfnamefont {L.}~\bibnamefont
  {Viola}}\ and\ \bibinfo {author} {\bibfnamefont {S.}~\bibnamefont {Lloyd}},\
  }\bibfield  {title} {\bibinfo {title} {Dynamical suppression of decoherence
  in two-state quantum systems},\ }\href
  {https://doi.org/10.1103/PhysRevA.58.2733} {\bibfield  {journal} {\bibinfo
  {journal} {Phys. Rev. A}\ }\textbf {\bibinfo {volume} {58}},\ \bibinfo
  {pages} {2733} (\bibinfo {year} {1998})}\BibitemShut {NoStop}%
\bibitem [{\citenamefont {Viola}\ \emph {et~al.}(1999)\citenamefont {Viola},
  \citenamefont {Knill},\ and\ \citenamefont {Lloyd}}]{LloydPRL1999}%
  \BibitemOpen
  \bibfield  {author} {\bibinfo {author} {\bibfnamefont {L.}~\bibnamefont
  {Viola}}, \bibinfo {author} {\bibfnamefont {E.}~\bibnamefont {Knill}},\ and\
  \bibinfo {author} {\bibfnamefont {S.}~\bibnamefont {Lloyd}},\ }\bibfield
  {title} {\bibinfo {title} {Dynamical decoupling of open quantum systems},\
  }\href {https://doi.org/10.1103/PhysRevLett.82.2417} {\bibfield  {journal}
  {\bibinfo  {journal} {Phys. Rev. Lett.}\ }\textbf {\bibinfo {volume} {82}},\
  \bibinfo {pages} {2417} (\bibinfo {year} {1999})}\BibitemShut {NoStop}%
\bibitem [{\citenamefont {Gordon}\ and\ \citenamefont
  {Kurizki}(2006)}]{Gordon2006}%
  \BibitemOpen
  \bibfield  {author} {\bibinfo {author} {\bibfnamefont {G.}~\bibnamefont
  {Gordon}}\ and\ \bibinfo {author} {\bibfnamefont {G.}~\bibnamefont
  {Kurizki}},\ }\bibfield  {title} {\bibinfo {title} {Preventing multipartite
  disentanglement by local modulations},\ }\href
  {https://doi.org/10.1103/PhysRevLett.97.110503} {\bibfield  {journal}
  {\bibinfo  {journal} {Phys. Rev. Lett.}\ }\textbf {\bibinfo {volume} {97}},\
  \bibinfo {pages} {110503} (\bibinfo {year} {2006})}\BibitemShut {NoStop}%
\bibitem [{\citenamefont {Yang}\ and\ \citenamefont {Liu}(2008)}]{Yang2008}%
  \BibitemOpen
  \bibfield  {author} {\bibinfo {author} {\bibfnamefont {W.}~\bibnamefont
  {Yang}}\ and\ \bibinfo {author} {\bibfnamefont {R.-B.}\ \bibnamefont {Liu}},\
  }\bibfield  {title} {\bibinfo {title} {Universality of uhrig dynamical
  decoupling for suppressing qubit pure dephasing and relaxation},\ }\href
  {https://doi.org/10.1103/PhysRevLett.101.180403} {\bibfield  {journal}
  {\bibinfo  {journal} {Phys. Rev. Lett.}\ }\textbf {\bibinfo {volume} {101}},\
  \bibinfo {pages} {180403} (\bibinfo {year} {2008})}\BibitemShut {NoStop}%
\bibitem [{\citenamefont {de~Lange}\ \emph {et~al.}(2010)\citenamefont
  {de~Lange}, \citenamefont {Wang}, \citenamefont {Rist\`{e}}, \citenamefont
  {Dobrovitski},\ and\ \citenamefont {Hanson}}]{HansonScience2010}%
  \BibitemOpen
  \bibfield  {author} {\bibinfo {author} {\bibfnamefont {G.}~\bibnamefont
  {de~Lange}}, \bibinfo {author} {\bibfnamefont {Z.~H.}\ \bibnamefont {Wang}},
  \bibinfo {author} {\bibfnamefont {D.}~\bibnamefont {Rist\`{e}}}, \bibinfo
  {author} {\bibfnamefont {V.~V.}\ \bibnamefont {Dobrovitski}},\ and\ \bibinfo
  {author} {\bibfnamefont {R.}~\bibnamefont {Hanson}},\ }\bibfield  {title}
  {\bibinfo {title} {Universal dynamical decoupling of a single solid-state
  spin from a spin bath},\ }\href {https://doi.org/10.1126/science.1192739}
  {\bibfield  {journal} {\bibinfo  {journal} {Science}\ }\textbf {\bibinfo
  {volume} {330}},\ \bibinfo {pages} {60} (\bibinfo {year} {2010})}\BibitemShut
  {NoStop}%
\bibitem [{\citenamefont {Khodjasteh}\ \emph {et~al.}(2011)\citenamefont
  {Khodjasteh}, \citenamefont {Dobrovitski},\ and\ \citenamefont
  {Viola}}]{KhodjastehPRA2011}%
  \BibitemOpen
  \bibfield  {author} {\bibinfo {author} {\bibfnamefont {K.}~\bibnamefont
  {Khodjasteh}}, \bibinfo {author} {\bibfnamefont {V.~V.}\ \bibnamefont
  {Dobrovitski}},\ and\ \bibinfo {author} {\bibfnamefont {L.}~\bibnamefont
  {Viola}},\ }\bibfield  {title} {\bibinfo {title} {Pointer states via
  engineered dissipation},\ }\href {https://doi.org/10.1103/PhysRevA.84.022336}
  {\bibfield  {journal} {\bibinfo  {journal} {Phys. Rev. A}\ }\textbf {\bibinfo
  {volume} {84}},\ \bibinfo {pages} {022336} (\bibinfo {year}
  {2011})}\BibitemShut {NoStop}%
\bibitem [{\citenamefont {Yang}\ \emph {et~al.}(2011)\citenamefont {Yang},
  \citenamefont {Wang},\ and\ \citenamefont {Liu}}]{LiuFrontiers2011}%
  \BibitemOpen
  \bibfield  {author} {\bibinfo {author} {\bibfnamefont {W.}~\bibnamefont
  {Yang}}, \bibinfo {author} {\bibfnamefont {Z.-Y.}\ \bibnamefont {Wang}},\
  and\ \bibinfo {author} {\bibfnamefont {R.-B.}\ \bibnamefont {Liu}},\
  }\bibfield  {title} {\bibinfo {title} {Preserving qubit coherence by
  dynamical decoupling},\ }\href {https://doi.org/10.1007/s11467-010-0113-8}
  {\bibfield  {journal} {\bibinfo  {journal} {Front. Phys.}\ }\textbf {\bibinfo
  {volume} {6}},\ \bibinfo {pages} {2} (\bibinfo {year} {2011})}\BibitemShut
  {NoStop}%
\bibitem [{\citenamefont {Tan}\ \emph {et~al.}(2014)\citenamefont {Tan},
  \citenamefont {Huang}, \citenamefont {Kuang},\ and\ \citenamefont
  {Wang}}]{Tan2013}%
  \BibitemOpen
  \bibfield  {author} {\bibinfo {author} {\bibfnamefont {Q.-S.}\ \bibnamefont
  {Tan}}, \bibinfo {author} {\bibfnamefont {Y.}~\bibnamefont {Huang}}, \bibinfo
  {author} {\bibfnamefont {L.-M.}\ \bibnamefont {Kuang}},\ and\ \bibinfo
  {author} {\bibfnamefont {X.}~\bibnamefont {Wang}},\ }\bibfield  {title}
  {\bibinfo {title} {Dephasing-assisted parameter estimation in the presence of
  dynamical decoupling},\ }\href {https://doi.org/10.1103/PhysRevA.89.063604}
  {\bibfield  {journal} {\bibinfo  {journal} {Phys. Rev. A}\ }\textbf {\bibinfo
  {volume} {89}},\ \bibinfo {pages} {063604} (\bibinfo {year}
  {2014})}\BibitemShut {NoStop}%
\bibitem [{\citenamefont {Pokharel}\ \emph {et~al.}(2018)\citenamefont
  {Pokharel}, \citenamefont {Anand}, \citenamefont {Fortman},\ and\
  \citenamefont {Lidar}}]{LidarPRL2018}%
  \BibitemOpen
  \bibfield  {author} {\bibinfo {author} {\bibfnamefont {B.}~\bibnamefont
  {Pokharel}}, \bibinfo {author} {\bibfnamefont {N.}~\bibnamefont {Anand}},
  \bibinfo {author} {\bibfnamefont {B.}~\bibnamefont {Fortman}},\ and\ \bibinfo
  {author} {\bibfnamefont {D.~A.}\ \bibnamefont {Lidar}},\ }\bibfield  {title}
  {\bibinfo {title} {Demonstration of fidelity improvement using dynamical
  decoupling with superconducting qubits},\ }\href
  {https://doi.org/10.1103/PhysRevLett.121.220502} {\bibfield  {journal}
  {\bibinfo  {journal} {Phys. Rev. Lett.}\ }\textbf {\bibinfo {volume} {121}},\
  \bibinfo {pages} {220502} (\bibinfo {year} {2018})}\BibitemShut {NoStop}%
\bibitem [{\citenamefont {Austin}\ \emph {et~al.}(2020)\citenamefont {Austin},
  \citenamefont {Zahid},\ and\ \citenamefont {Chaudhry}}]{Austin2020}%
  \BibitemOpen
  \bibfield  {author} {\bibinfo {author} {\bibfnamefont {S.}~\bibnamefont
  {Austin}}, \bibinfo {author} {\bibfnamefont {S.}~\bibnamefont {Zahid}},\ and\
  \bibinfo {author} {\bibfnamefont {A.~Z.}\ \bibnamefont {Chaudhry}},\
  }\bibfield  {title} {\bibinfo {title} {Geometric phase corrected by initial
  system-environment correlations},\ }\href@noop {} {\bibfield  {journal}
  {\bibinfo  {journal} {Phys. Rev. A}\ }\textbf {\bibinfo {volume} {101}},\
  \bibinfo {pages} {022114} (\bibinfo {year} {2020})}\BibitemShut {NoStop}%
\bibitem [{\citenamefont {M\"{u}ller}\ \emph {et~al.}(2018)\citenamefont
  {M\"{u}ller}, \citenamefont {Gherardini},\ and\ \citenamefont
  {Caruso}}]{MullerSciRep2018}%
  \BibitemOpen
  \bibfield  {author} {\bibinfo {author} {\bibfnamefont {M.}~\bibnamefont
  {M\"{u}ller}}, \bibinfo {author} {\bibfnamefont {S.}~\bibnamefont
  {Gherardini}},\ and\ \bibinfo {author} {\bibfnamefont {F.}~\bibnamefont
  {Caruso}},\ }\bibfield  {title} {\bibinfo {title} {Quantum parameter
  estimation in a dissipative environment},\ }\href
  {https://doi.org/10.1038/s41598-018-32434-x} {\bibfield  {journal} {\bibinfo
  {journal} {Sci. Rep.}\ }\textbf {\bibinfo {volume} {8}},\ \bibinfo {pages}
  {14278} (\bibinfo {year} {2018})}\BibitemShut {NoStop}%
\end{thebibliography}
\end{document}